\documentclass[12pt,preprint]{aastex}
\usepackage{graphicx}

\begin{document}
\slugcomment{Submitted to ApJ}

\title{Discovery of an eclipsing X-ray binary with a 32.69 hour period in M101:
an analog of Her X-1 or LMC X-4?}

\author{Ji-Feng Liu\altaffilmark{1}, Rosanne Di Stefano\altaffilmark{1},
Jeffrey McClintock\altaffilmark{1}, Albert Kong\altaffilmark{2},Joel
Bregman\altaffilmark{3},and Kip Kuntz\altaffilmark{4}}

\altaffiltext{1}{Harvard-Smithsonian Center for Astrophysics}
\altaffiltext{2}{Massachusetts Institute of Technology}
\altaffiltext{3}{University of Michigan}
\altaffiltext{4}{JHU/GSFC}

\begin{abstract}

We report the discovery of an eclipsing X-ray binary in M101, the first such
system to be discovered outside the Local Group.  Based on a sequence of 25
Chandra observations that sample a wide range of orbital phases, we find a
period of 32.688 $\pm$ 0.002 hours, which we interpret as an orbital period.
The folded light curve exhibits an eclipse lasting about 8 hours, suggesting a
compact orbit in an nearly edge-on configuration.  The X-ray binary has an
average luminosity of $L_X(0.3-8 {\rm keV}) \approx 1.3\times10^{38}$ erg/sec,
with only one out of the 25 observations significantly lower in flux than the
average light curve.The presence of the eclipse and the $\sim$ 1.4-day orbital
period suggests that this source is an analog of the well studied eclipsing
X-ray binary pulsars Her X-1 or LMC X-4.  Combining the Chandra data and the
HST ACS/WFC images, we have identified several possible optical counterparts,
including an O5-O3 star with V = 25.0.  Follow-up optical monitoring
observations should be able to identify the donor and further constrain the
orbital properties.  

\end{abstract}


\section{INTRODUCTION}

The detections of X-ray binaries have recently begun in galaxies outside the
Local Group with the advent of the Chandra X-ray observatory.
Owing to its unprecedented sensitivity and spatial resolution, it is possible
to resolve individual X-ray point sources in these galaxies.
In the past six years, Chandra has studied about one third ($\sim$70/220) of
the galaxies with $D_{25}>1^\prime$ within 10 Mpc (Liu 2005); these galaxies
span the whole spectrum of  galaxy types, including spirals (e.g., M81, Tennant
et al.  2001; M51, Terashima \& Wilson 2004), ellipticals (e.g., NGC1399,
Angelini et al.  2001; NGC4697, Sarazin et al. 2000), starburst galaxies (e.g.,
Antenna and M82, Zezas et al. 2000), and colliding galaxies (NGC7714/7715,
Smith et al. 2005).
The distributions and luminosity functions of the X-ray binaries in these
galaxies, as compared to those in the Milky Way (Grimm et al. 2002) and M31
(Kong et al. 2003), show significant dependence on the galaxy type and star
formation properties.

An ultra-deep Chandra ACIS survey (PI: Kuntz) has been executed for the face-on
spiral M101 (NGC5457) at a distance of 6.8 Mpc (Freedman et al. 2001).
Twenty four Chandra observations were tiled to achieve a full coverage of M101, with a
cumulative exposure of one mega second within 4$^\prime$ of the nucleus.
Such a deep exposure can reach a detection limit of 4$\times$$10^{35}$ erg/sec,
only  a factor of two more luminous than current Chandra studies of M31 (Kong
et al.  2002).
This survey allows detection of most low-mass X-ray binaries, most high-mass
X-ray binaries, most young supernova remnants and many middle-aged ones,
super-bubbles and giant HII regions.
The observations were carried out over the year from January 2004 to January
2005, allowing studies of the variability and transient behaviors of these
sources.

Here we report the discovery of a 32.688$\pm$0.002 hour period for an eclipsing
X-ray binary from the M101 ultra-deep survey and an early observation in 2000.
The Chandra observations and the search for periodicities are described in
section 2.
Additional optical observations with HST ACS are described in Section 3. In
Section 4, we describe the constraints the 32.69 hour period places on the
accretor and donor masses, and discuss the prospects to monitor and identify
the donor at optical wavelengths.
This X-ray source has been detected previously in ROSAT HRI and PSPC
observations and named as M101-X7 (Liu \& Bregman 2005; H36/P21 in Wang, Immer
\& Pietsch 1999), but the periodicity could not have been detected given the
low photon counts and the small number of observations.


\section{Chandra Observations}

M101-X7 was observed in 25 Chandra observations spanning five years (Table 1),
including 24 observations from the M101 ultra-deep survey and an earlier
observation (ObsID 934, 2000-03-26).
All the observations were downloaded from the Chandra Data Archive,
and processed with CIAO 3.3.0.1 and CALDB 3.2.1.
WAVDETECT was run with the scale set of ($0\farcs5$, $1\farcs0$, $2\farcs0$,
$4\farcs0$) to detect point sources on the X-ray images.

The source was detected in 25 observations at an average location of
R.A.=14:03:36.0, Decl.=54:19:25, thus named as CXO J140336.0+541925 following
the CXO naming convention.
The source was at large off-axis angles in most observations, and was elongated
at different elongation angles due to the different positioning of the aim
points.
The merged profile from all observations was symmetric and could be fitted by a
Gaussian with FWHM of $1\farcs7$, consistent with the point source diameter
($1\farcs7$) for enclosed energy fraction of 50\% at an off-axis angle of 5
arcminutes.

The light curve was constructed to demonstrate variations during individual
observations.
The photons (0.3 keV $<$E$<$ 8 keV) were extracted from the source region
enclosing 95\% of the total photons reported by wavdetect. 
The time bin was adjusted to have an average of $\ge$15 counts per bin and no
more than 128 bins.
Among the 25 light curves, emphases were put on a group of 18 ``good'' light
curves for the 18 observations with more than 100 photons, with 17 observed in
2004 during the M101 ultra-deep survey, and one (ObsID 934) observed in early
2000.
Among the 18 ``good'' light curves, 12 light curves showed apparent variations
suggestive of periodicities, with apparent minima in seven of them (Table 1).

Two techniques were applied to search for possible periods.
The first was the Lomb-Scargle method devised for unevenly spaced data (Lomb
1976; Scargle 1982).
The Lomb periodogram was computed for the bins in the 18 ``good'' light curves
(Figure 1), which showed significant peaks at 32.69 hours, 22.44 hours, 41.75
hours, and 16.10 hours.
The probabilities for these peaks to originate from random fluctuations of
photons in the absence of true signals are less than $10^{-10}$, suggestive
of the presence of true signals.
This, however, does not mean all peaks in the Lomb periodogram are true
periods.  Indeed, some peaks may be just period aliases.
Another method applied to the data was the phase dispersion minimization method
as demonstrated in Stellingwerf (1978).
A rough search of periods in the range of 5 hours to 50 hours with steps of
0.045 hours revealed significant phase dispersion ($\Theta$) minima at 32.72
hours, 44.87 hours, 33.53 hours, 31.91 hours, and 22.82 hours (Figure 2).
Searches were then carried out with finer steps of $2\times10^{-4}$ hours
around these minima as illustrated by the inserts in Figure 2. 

The above searches resulted in a collection of period candidates as peaks in
the Lomb periodogram and/or phase dispersion minima.
The light curves were folded with each candidate period to compute an average
light curve, which was then  compared to the individual light curves to check
for phase preservation. 
Only for the period of 32.69 hours were all 25 individual light curves
consistent with the average light curve, while for other periods some
individual light curves were offset from the average light curve by more than
$90^\circ$.
We conclude that the true period is 32.69 hours. 
To understand the origin of the period aliases, we simulated data using the
period of 32.69 hours, with the average light curve described below and  the
time windows for the observations.
The simulated data were able to reproduce the period aliases, suggesting they
can be attributed to the time windows for the observations and the shape of the
underlying light curve.


The average light curve for the period of 32.69 hours, computed with 16 light
curves observed in the ultra-deep survey, showed a clear eclipse at phase 0.4
and a minor dip at phase 0.95 (Figure 3).
The average light curve was compared to all individual light curves, including
the 18 ``good'' light curves with more than 100 photons and the other seven
light curves with fewer than 100 photons (Figure 4).
Remarkable consistency in the phase and the flux level was found between the
average and all individual light curves, except for the light curve of ObsID
5340 which was significantly lower than the average.
Deep eclipses of about eight hours were seen in a number of observations, with
steep ingress and egress both occurring within one time bin ($\sim$1 hour).
We thus estimate the eclipse duration to be 8 $\pm$ 1 hours, corresponding to
an eclipse full angle of $88^\circ \pm 11^\circ$.
Residual emission in the eclipse was present in some individual light curves
and in the average light curve.
The apparently slow ingress and egress of the eclipse in the average light
curve were artifacts due to the averaging of 16 light curves, all of which were
not aligned exactly.


The period estimate can be improved by using the light curve in ObsID 934,
which extended the observation time line from one year ($\sim268$ cycles) to
five years ($\sim1279$ cycles).
For the observations in 2004, the times of the seven minima can be determined
to better than $30^\circ$ relative to the average light curve, leading to a
fractional error in period of $\Delta P/P\sim (30^\circ/360^\circ)/268 \sim
0.0003$.
A slight change in the period does not change the phases much for observations
in 2004, but leads to much larger phase changes for the observation ObsID 934.
By fine tuning the period, the light curve of ObsID 934 was found to align up
with the average light curve to better than $30^\circ$ at $P=32.688$ hours
(Figure 5). 
The fractional error in the period is $\sim(30^\circ/360^\circ)/1279\sim
6.5\times10^{-5}$, equivalent to an error of 0.002 hours or 7.7 seconds.
Thus, the period is 32.688 $\pm$ 0.002 hours, and throughout this paper we use
the following ephemeris: $\phi = 0.0 $ for JD 2450814.49927 + n*(1.36200 $\pm$
0.00008).

The source spectrum was extracted for each observation yielding more than 300
counts.
Six spectra were extracted and fitted by absorbed power-law models (Table 1).
The model parameters were poorly constrained, and were consistent with each
other except for the spectrum in ObsID 934.
The five similar spectra were combined for fitting to better constrain the
parameters. This led to $n_H/10^{21} = 3.3\pm0.5$ and $\Gamma = 3.4\pm0.3$ for
the absorbed power-law model ($\chi_\nu^2/dof = 1.06/50$), and $n_H/10^{20} =
9.6\pm2.6$ and $T_{in} = 0.45\pm0.4$ keV for the absorbed multi-color disk
model ($\chi_\nu^2/dof = 1.25/50$).
The count rates were converted to unabsorbed fluxes in 0.3-8 keV with the conversion
factor 1 count/ksec = 4.86 $\times 10^{-15}$ erg/sec/cm$^2$ (i.e. 2.7 $\times
10^{37}$ erg/sec at a distance of 6.8 Mpc) derived from the combined power-law
fit.
The unabsorbed luminosities were in the range of 0.2-2.3 $\times10^{38}$
erg/sec with an overall average of 1.3 $\times10^{38}$ erg/sec (Table 1).
If the combined multi-color disk model was used, the conversion factor would be
1 count/ksec = 4.33 $\times 10^{-15}$ erg/sec/cm$^2$ (i.e. $2.4\times10^{37}$
erg/sec), and the average luminosity was 1.2 $\times10^{38}$ erg/sec.

A significant soft excess below 0.5 keV was present in the spectrum in ObsID
934. 
This led to a much lower $n_H$ value (still higher than the Galactic value $n_H
= 1.6 \times 10^{20}$ cm$^{-2}$) in the absorbed power law model fit.
The spectrum can be fitted with an absorbed power law plus multi-color disk
model, with $n_H/10^{21} = 1.9\pm0.7$, $\Gamma = 3.0\pm0.3$ and $T_{in} = 64
\pm 11$ eV ($\chi_\nu^2/dof = 1.084/19$). 
The spectral fit was slightly improved as compared to the one component models,
implying possible presence of a $\sim60$eV soft component in 2000 that was
missing from subsequent observations.
However, five other sources, randomly distributed over the chip in ObsID 934,
exhibited similar soft excesses that were absent in subsequent observations,
suggesting a possible alternative origin attributable to the uncertainties in
the calibration of the instrumental soft response in the early observations.
%





The spectrum is expected to become harder (softer) at the eclipse ingress
(egress) due to additional absorption by the blocking secondary, and the
residual emission in the eclipse, presumably from the stellar winds, is
expected to be very soft.
While the small number of photons during the eclipse does not allow a detailed
study of the additional absorption and stellar winds, we can at least test
whether there is additional absorption and/or stellar winds thus hardening
and/or softening of the spectrum.
The phase for each photon from all observations was computed for the period of
32.688 hours.
The 340 photons with phases 0.25-0.50 were considered in the eclipse, and the
4490 photons with phases 0.50-1.25 were considered outside the eclipse.  
The ``raw'' spectra for the two groups were plotted in Figure 6.
Compared to the ``non-eclipse'' spectrum, the ``eclipse'' spectrum exhibited a
lack of soft photons below 2 keV and an excess of hard photons above 4 keV.
A Kolmogorov-Smirnov test showed the two spectra were the same at a 0.02 level,
indicating the hardening of the ``eclipse'' spectrum was true at a 98\%
confidence level.
To reveal spectral changes associated with ingress and egress of the eclipse,
we calculated the hardness ratio as (H-S)/(H+S) for phase bins of 0.04, as
plotted in Figure 7 for three sets of soft and hard bands. 
The softening was present at the minor dip ($\phi = 0.9$) in all three band
sets, but with less than $1-\sigma$ significance.
Slight hardening was present at ingress ($\phi=$ 0.25) for band sets
S(0.1-1)/H(1-8) and S(0.1-1.3)/H(1.3-8), however, it was absent at ingress for
band set S(0.1-0.7)/H(0.7-8), and the reverse softening is absent at egress for
all band sets.
Indeed, the hardness ratios for most phase  bins were consistent with the
average hardness ratios within $1-\sigma$ error bars (Figure 7).

\section{Optical Observations}

M101-X7 was observed with HST ACS/WFC in the F658N ($H_\alpha$; 2440 seconds)
filter on 2004-02-10, and the F435W (B; 900 seconds), F555W (V; 720 seconds),
and F814W (I; 720 seconds) filters on 2002-11-16. 
The images were downloaded from the STSci archive and calibrated on the fly
with the best available calibration files.
The drizzled images with geometric distortion corrected and cosmic rays removed
were used to compare with the X-ray image from ObsID 6115, in which X7 was
off-axis by $\sim$$2^\prime$.
X7 was registered onto the optical images with the help of a nearby source, CXO
J140339.3+541827, which was positionally coincident with an extremely red
bright extended object in a sparse stellar field.
The positional uncertainty of X7 on the optical images comes from the
centroiding errors of X7 and CXO J140339.3+541827 and the ACIS plate scale
variation, and is about $0\farcs3$.
Due to its large off-axis angle, the X-ray emission of X7 was dispersed to
within an ellipse of $1\farcs4 \times 1\farcs2$ as overplotted on the optical
images in Figure 8.


X7 is located in the center of a super-bubble between the spiral arms as shown
on the $H_\alpha$ image.
This super-bubble shows shell structures with a size of $\sim
8^{\prime\prime}\times8^{\prime\prime}$ (i.e., $\sim 270pc \times270pc$ at 6.8
Mpc). The shells exhibit non-concentric irregularities, indicating multiple
supernova explosions at different off-center locations.
Several possible counterparts were found within the overplotted ellipse in
Figure 8, with the brightest labeled from `A' to `J'.
To detect point sources and compute the photometry, the PSF-fitting package
DOLPHOT (Dolphin 2000) was run on these images, which resolved some of the
labeled counterparts into several stars.
These counterpart candidates have V = 23.5 - 29.0 mag as listed in Table 2.
The V-I colors range from -0.7 to 3.3 mag, with an average of 0.7 mag.
Such colors indicate the presence of heavy reddening commonly seen in star
forming regions, or stellar spectral types later than G0 (V-I = 0.7 for G0V,
V-I = 3.4 for M7V). At the distance of M101, dwarf stars later than G0V cannot
be detected, but G/M supergiants can be detected.

The counterpart candidate A1 was closest to the nominal X-ray position, and had
V-I = $0.2\pm0.11$ mag and V = $24.96 \pm 0.05$ mag, suggestive of an B0-O9V
star ($M_V = -4.0 \sim -4.5$ mag) or brighter at 6.8 Mpc ($\mu=29.2$ mag) in
M101 given the absolute magnitude.
To search for the spectral types with consistent extinction and reddening, we
first calculated the extinction $A_V$ for each spectral spectral type above
B0V, then the reddening E(B-V) $= A_V/R_V = A_V/3.1$ and E(V-I) assuming the
standard extinction law (Cardelli, Clayton \& Mathis, 1989).
The search pointed to spectral types of O5-O3 ($M_V = -5.7 - -6.0$ mag, V-I =
-0.47 mag).  For such stars, the extinction $A_V$ = 1.5 - 1.8 mag, leading to
the reddening E(B-V) = 0.48 - 0.58 mag and E(V-I) = 0.57 - 0.69 mag, consistent
with the observed E(V-I) = 0.67 $\pm$ 0.11 mag.
This reddening corresponded to $n_H = 2.9 - 3.5 \times 10^{21}$ cm$^{-2}$
(Bohlin, Savage, \& Drake, 1978), consistent with the $n_H$ value derived from
the power-law fit to the combined X-ray spectra.


As a side note, M101-X7 has drawn much attention due to its positional
coincidence with a supernova remnant candidate MF83 (Matonick \& Fesen 1997). 
Early ROSAT observations did not have enough photons and the spatial resolution
to determine whether X7 was a point source or a diffuse source.  If X7 were diffuse
emission from shock heated gas in the remnant, its high luminosity would
require an explosion energy two orders of magnitude higher than those of normal
supernova and make it a hypernova (Wang 1999). 
Subsequent ground-based observations showed that MF83 is a star formation
region with several HII regions and a large ionized shell that is a super-bubble
based on its physical properties (Lai et al. 2001). 
If the X-ray emission were indeed diffuse, this shell would be an X-ray bright
super-bubble whose thermal energy would require either a hypernova or 10-100
supernova in the past $10^6$ years.
The recent Chandra observations, however, proved that X7 is consistent with a
point-like source in shape. Furthermore, the presence of periodicity and eclipse
in the X-ray light curves provides unmistaken evidence for the point-like nature
of X7.

\section{Discussion}

M101 X7 (CXO J140336.0+541925)  was observed in 25 Chandra observations
spanning five years, and exhibited significant modulations in twelve
observations, with apparent minima present in seven observations.
The Lomb periodogram and the phase dispersion minimization were applied in a
search for periodicities, revealing a period of 32.688 $\pm$ 0.002 hours for
this source.
The average light curve folded with this period shows a deep eclipse with an
eclipse full angle of $\theta_e \sim 88^\circ \pm 11^\circ$, suggesting that
this source is an eclipsing X-ray binary in a nearly edge-on configuration.
At a distance of 6.8 Mpc, this is the first eclipsing X-ray binary discovered
outside the Local Group. 
This discovery can only have been made with the superb sensitivity and spatial
resolution of the Chandra X-ray Observatory, and with the M101 ultra-deep
sequence of observations sampling different orbital phases.

Residual emission was present in the eclipse, indicating that the X-rays are
scattered or otherwise processed by, e.g., the strong wind as in the high mass
X-ray binaries Cen X-3 (e.g., Schreier et al. 1972) and Vela X-1 (Becker et al.
1978).
The presence of strong winds may lead to a slight over-estimate of the eclipse
full angle, and its possible variability may have contributed to the
mis-alignment of the ingress and egress in some observations.
The residual emission from stellar winds will make the spectrum in the eclipse
appear softer, while the additional absorption by the blocking secondary will
make it harder.
A preliminary study, limited by the small number of photons, showed that the
hardness of the spectrum did not change significantly with the orbital phases,
except for slight hardening present in the eclipse.

Much information can be inferred for this binary from the light curves.
The edge-on viewing geometry as suggested by the presence of eclipse in the
light curve renders inapplicable the beaming mechanism which enhances the
radiation along the polar axis. Thus, its average luminosity of $\bar
L_X$=1.3$\times10^{38}$ erg/sec constrains the accretor mass through the
Eddington limit to be larger than 1 $M_\odot$.
Despite the differences in the average fluxes in individual observations, 24
out of the 25 light curves were roughly consistent with the average light
curve, and only one (i.e., 4\%) was significantly lower in flux than the
average light curve.
This is distinctly different from Be X-ray binaries with typical periods of
100-300 days, which are in high flux states only briefly when the accretors are
near the perigee of their highly elliptical orbits.
The relatively shorter period, the wide eclipse, and the nearly uniform
distribution of the high flux states suggest that this binary has a much more
compact and highly circularized orbit.

The presence of the 32.69 hour period and the X-ray eclipse place strong
constraints on the binary properties if we assume the period is the orbital
period and the donor fills its Roche lobe.
The period $P$ can be computed with Kepler's law $P = 2\pi
a^{3/2}/\sqrt{G(M_1+M_2)}$, in which the separation $a$ can be related to the
Roche lobe radius $R_{cr}$ by $R_{cr} = a\cdot f(q) = a \cdot 0.49 q^{2/3}/[0.6
q^{2/3} + ln(1+q^{1/3})]$ (Eggleton 1983).  Here $M_1$ is the accretor mass,
$M_2$ is the donor mass, and $q = M_2/M_1$ is the mass ratio.
If the donor is filling its Roche lobe, i.e., $R_2 = R_{cr}$, the period will
be determined solely by the accretor mass and the donor mass and radius that
can be determined empirically by the donor's spectral type.
A discrete set of donors of the spectral types as tabulated in {\it the
Astrophysical Quantities} (Allen 2000) are tried to obtain this period for an
accretor mass range 1-10$^4$ $M_\odot$.
This period can only be obtained for some dwarf stars; all giants and
supergiants give much longer periods.

The donor star can have evolved off the main sequence even if it is not yet a
giant/supergiant star.
To consider such donors, we first considered main sequence dwarf donors
spanning a continuous mass range of 0.5-120 $M_\odot$, with radii $R_2 = r_0$
interpolated from the empirical mass-radius relation based on dwarf stars of
all spectral types.
We then introduced a helium core mass $M_c$ for each star, using a standard
relationship between core mass and stellar radius: $R_2 = r_0 + 3700 M_c^4 /(1+
M_c^3 + 1.75 M_c^4)$.
In the donor mass-accretor mass phase plane the viable binary systems with
donors of $M_c=0$ are shown as the thickest line, while the viable binary
systems with donors of $M_c>0$ populate the regions below and to the left of
this line (Figure 9).
The maximum eclipse full angles $\theta_E$ are computed for the viable binary
systems with $sin(\theta_E/2) = R_2/a$ assuming the inclination angle
$i=90^\circ$, and the observed large $\theta_e$ ($=\theta_E sini$) favors those
systems with massive donors (Figure 9). Such systems have a total mass $\le
10^2 M_\odot$, and the radii of the orbits are less than 1 light minute,
rendering barycenter corrections unnecessary.


The accretion rates are calculated for the viable binary systems assuming
donors overflow the Roche lobe under conservative mass transfer.
Invoking conservation of total angular momentum, the accretion rate can be
expressed as $\dot M_2/M_2 = N/D$. Here the denominator can be expressed as
$4[1-(M_c/\bar M_c)^2] + 2(1-q) - (1+q) dlnf/dlnq$ for conservative mass
transfer, and the numerator $N$ can be approximated as ${2.3\times10^{-7} \over
R_2 } [{(M_1+M_2)^2 \over M_1} {R_2^4 \over a^5}]$ when magnetic braking
dominates the orbital angular momentum loss.
Calculations show that the accretion rate decreases for smaller mass ratio
$q$'s and larger $M_c$'s; the contours are plotted in Figure 9 for $\dot M_2$ =
1, 0.1, 0.01, $10^{-3}$, $10^{-4}$ and $10^{-5}$ in unit of $\bar L_X/c^2$.
For a region with $q>1$ to the right of $\dot M_2$ = $\bar L_X/c^2$, $D\le0$,
indicating the breakdown of the adopted expression and very large mass transfer
rates that likely lead to common envelopes unless the accreted materials can be
dissipated effectively.

Similar orbital periods and eclipses have been detected in two well studied
X-ray binaries, i.e., the intermediate-mass X-ray binary Her X-1 and the
high-mass X-ray binary LMC X-4 as shown in Figure 9.
Her X-1 is an eclipsing binary pulsar with a pulse period of 1.24 seconds and
an orbital period of 1.7 days discovered by Uhuru (Tananbaum et al. 1972).
Its companion, HZ Her, is a $\sim2M_\odot$ A/F star (Bahcall \& Bahcall 1972)
filling its Roche lobe in a compact circular orbit.
The X-ray light curve for Her X-1 consists of alternating main on-state,
off-states, and short on-states with a super-orbital period of 35 days, which
is believed to be caused by the precession of a tilted accretion disk, and has
been studied by almost all X-ray missions (cf, Shakura et al.  1998).
LMC X-4 was first discovered by Uhuru (Giacconi et al. 1972), and was
identified with an O7 III-V star exhibiting a binary period of 1.4 days
(Chevalier \& Ilovaisky 1977).
The X-ray pulsations were discovered with a period of 13.5 seconds, the Doppler
variations of which revealed the system as a neutron star in circular orbit
with a 17 $M_\odot$ donor nearly filling its Roche lobe (Kelley et al. 1983).
The X-ray light curve for LMC X-4 exhibits eclipses, occasional flares, and a
30.5 day super-orbital modulation as prototyped by Her X-1 (Lang et al. 1981).

M101 X7 may be an M101 analog for Her X-1 or LMC X-4 given the similarities in
the orbital period and the presence of eclipses.
There is, however, no evidence for a super-orbital period of about 30 days for
M101 X7, and the pulse period cannot be detected given the low count rate of
about 0.005 cnt/sec.
Furthermore, its duty cycle for the low flux state is $\sim$ 4\%, while the duty
cycle for the low flux state (off-state) is $\ge$50\% in Her X-1 and LMC X-4.
In addition, its eclipse full angle is much larger than those for Her X-1 and
LMC X-4 (both $\sim50^\circ$), suggesting a higher inclination angle and/or a
more massive donor than for Her X-1 or even LMC X-4 (Figure 9).
Note, however, that the uncertainties in the duty cycle and eclipse angle are
large, given the small number of photons collected and the temporal sampling.
This makes direct comparison with Her X-1 and LMC X-4 difficult.

The donor and accretor masses can be further constrained by monitoring and
identifying the donor in the optical.
Our analyses of the HST ACS/WFC archive observations have resulted in several
candidates for the donor, which span a large range of magnitude and mass.
The donor mass, thus the primary mass, can be further constrained if the donor
can be identified with a single counterpart.
For example, were the O5-O3 star A1 (60-120 $M_\odot$) identified as the
secondary, the binary system would be constrained to the lower right corner in
Figure 9, with the primary mass lower than 10 $M_\odot$.
This identification can be achieved by photometric monitoring of these
counterparts over an orbital period, because the optical light curve should
exhibit eclipse and ellipsoidal modulations given the edge-on viewing geometry
of the binary.
These observations can be carried out with either HST or large ground-based
telescopes with adaptive optics such as Gemini-North in the north hemisphere.

\acknowledgements

We are grateful for the service of Chandra Data Archive. We would like to thank
Francis A. Primini, Saku Vrtilek, Harvey Tananbaum, Pavlos Protopapas, Paul P.
Plucinsky and Richard J. Edgar for helpful discussions. We would like to thank
XXX (the referee) for constructive suggestions. JNB  would like to thank NASA
for support from grant NAG5-10765.




\begin{figure}
\plotone{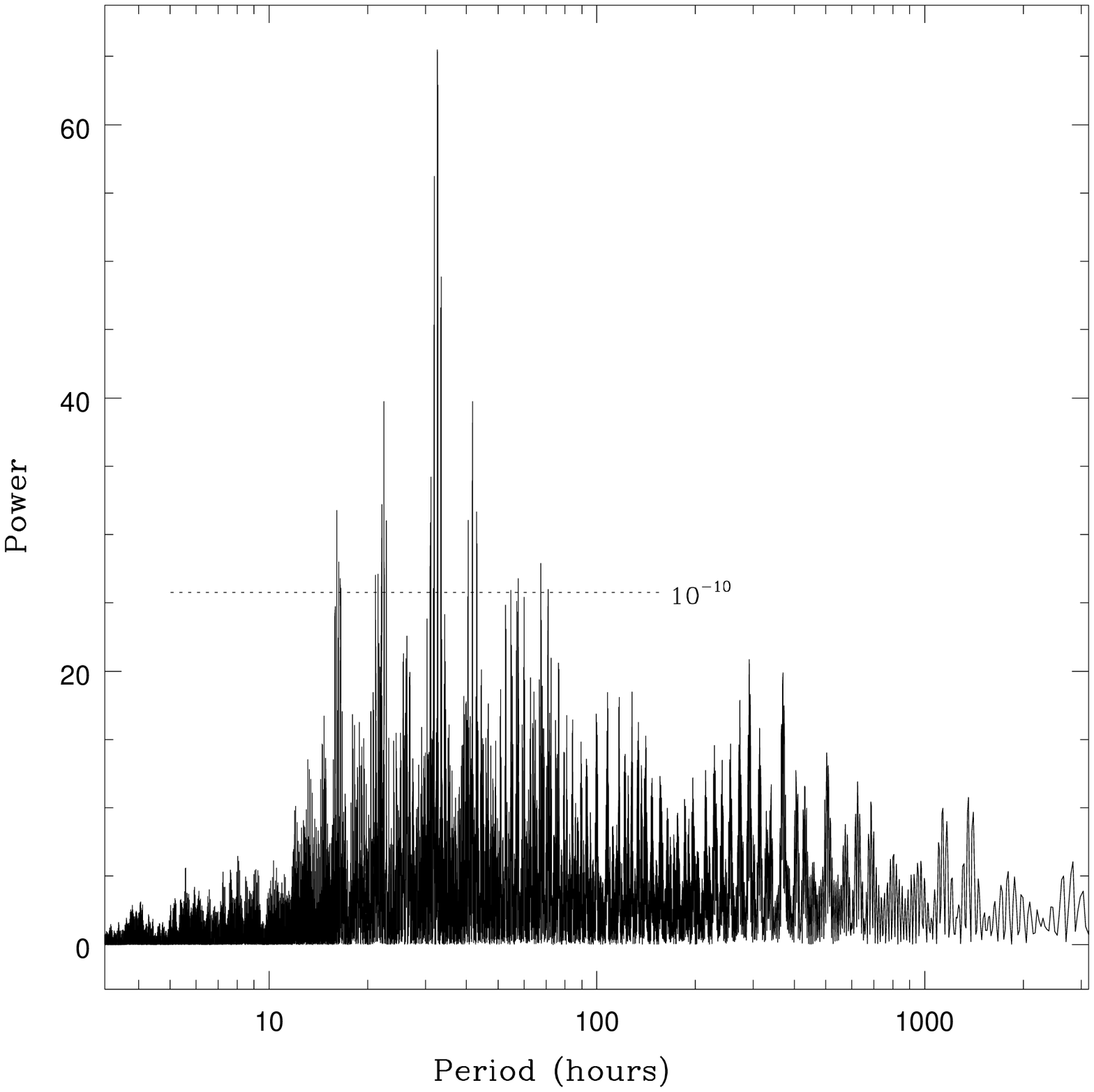}
\caption{Search for periodicities  for M101-X7 with the
Lomb-Scargle method.  The periodogram was computed for the 18 ``good'' light
curves.  The power 25.8 corresponds to the probability of $10^{-10}$ for the
periods to originate from random fluctuations in the absence of true 
signals.  }

\end{figure}


\begin{figure}
\plotone{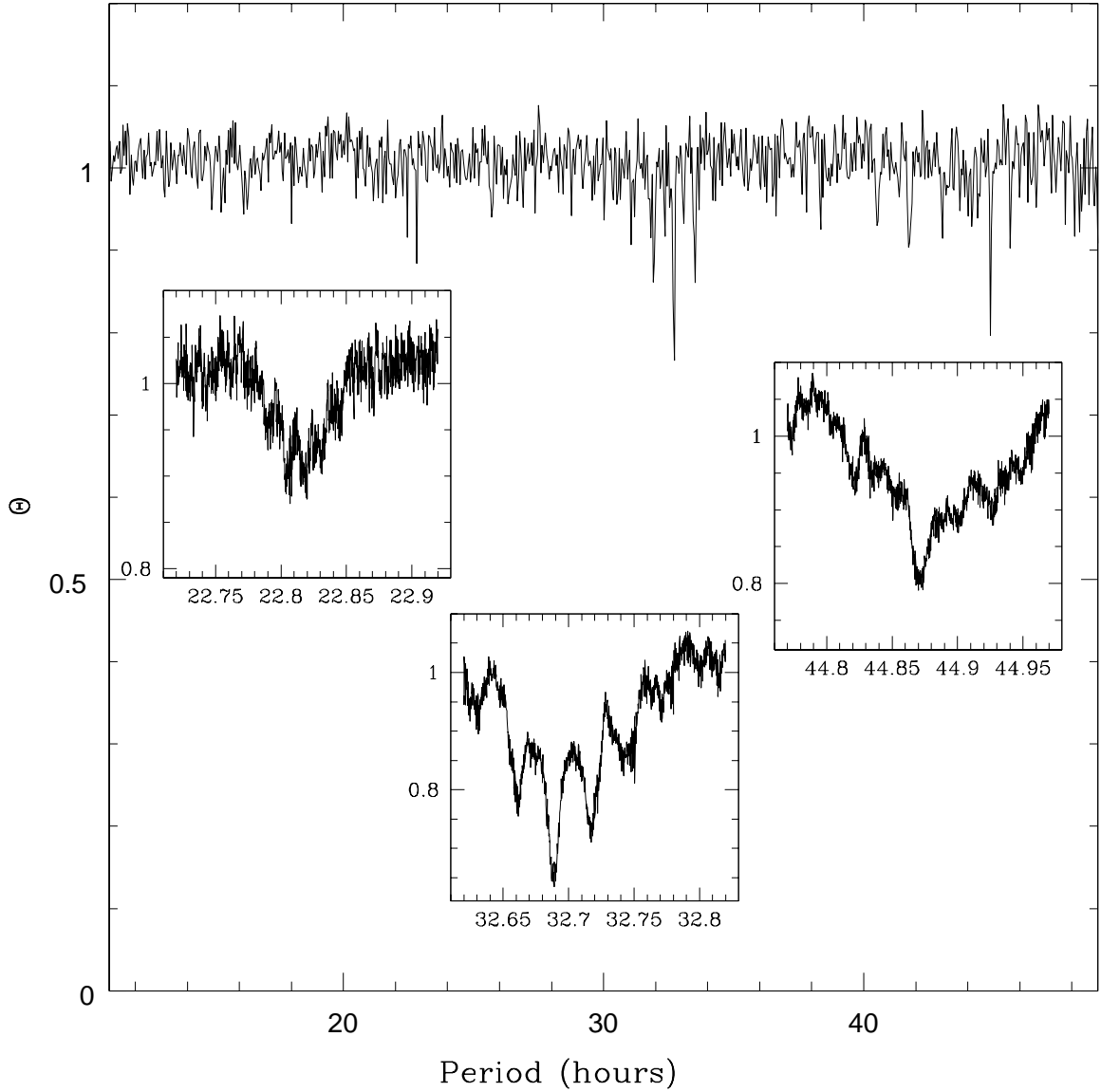}
\caption{Search for periodicities for M101-X7 with the phase dispersion
minimization method.  The phase dispersion was calculated for the 18 ``good''
light curves, first with trial period steps of 0.045 hours, then with finer
steps of $2\times10^{-4}$ hours around the local minima. The three inserts show
the finer searches around 22.82 hours, 32.72 hours, and 44.87 hours as
examples. }

\end{figure}


\begin{figure}
\plotone{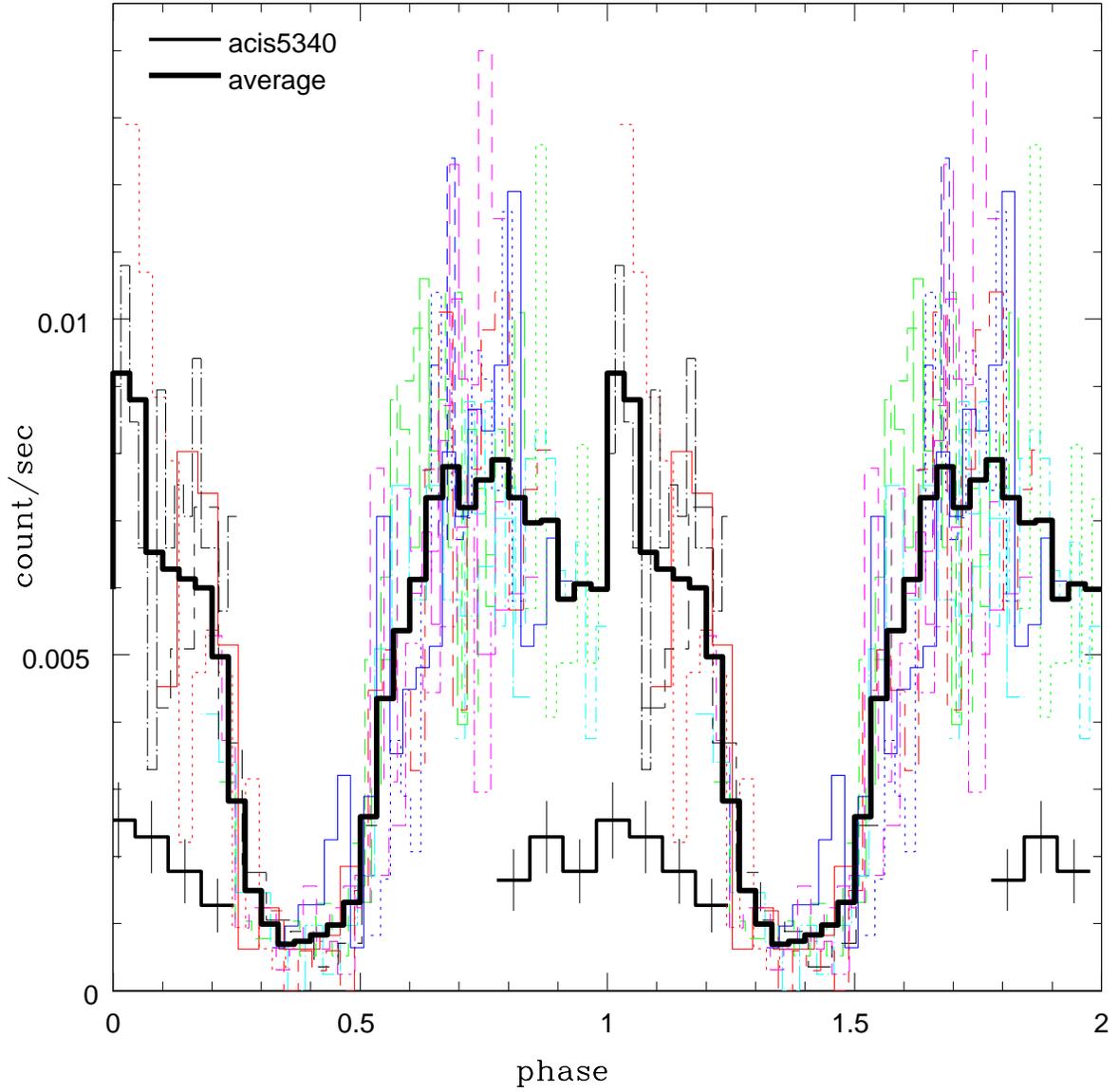}

\caption{The light curves folded with the period of 32.69 hours for the 17
observations in the M101 ultra-deep Chandra survey. The thickest histogram is
the average light curve from 16 light curves excluding ObsID 5340, which is
significantly lower than the average light curve. }

\end{figure}


\begin{figure} 
{\includegraphics[width=148pt,height=139pt]{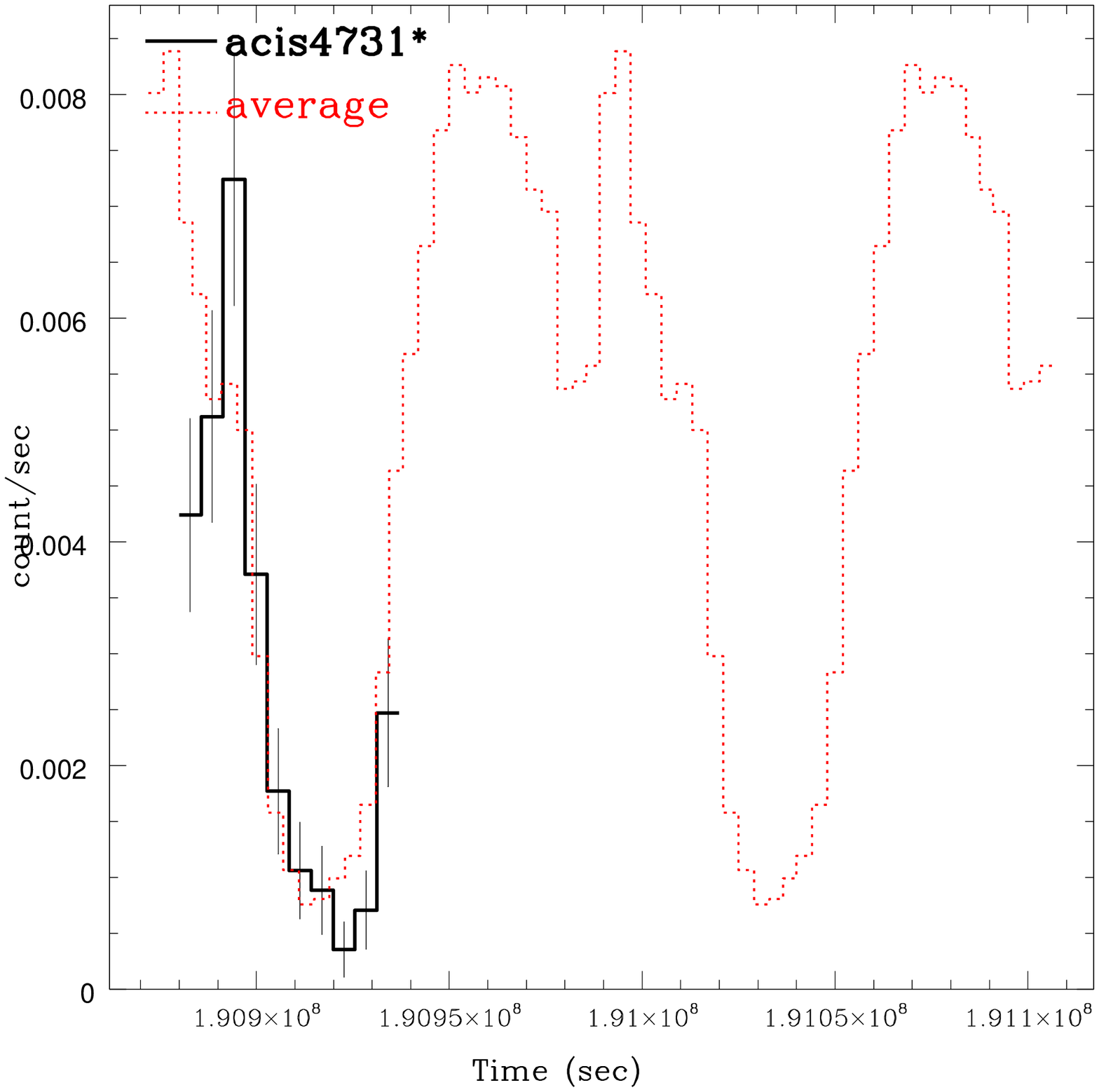}}
{\includegraphics[width=148pt,height=139pt]{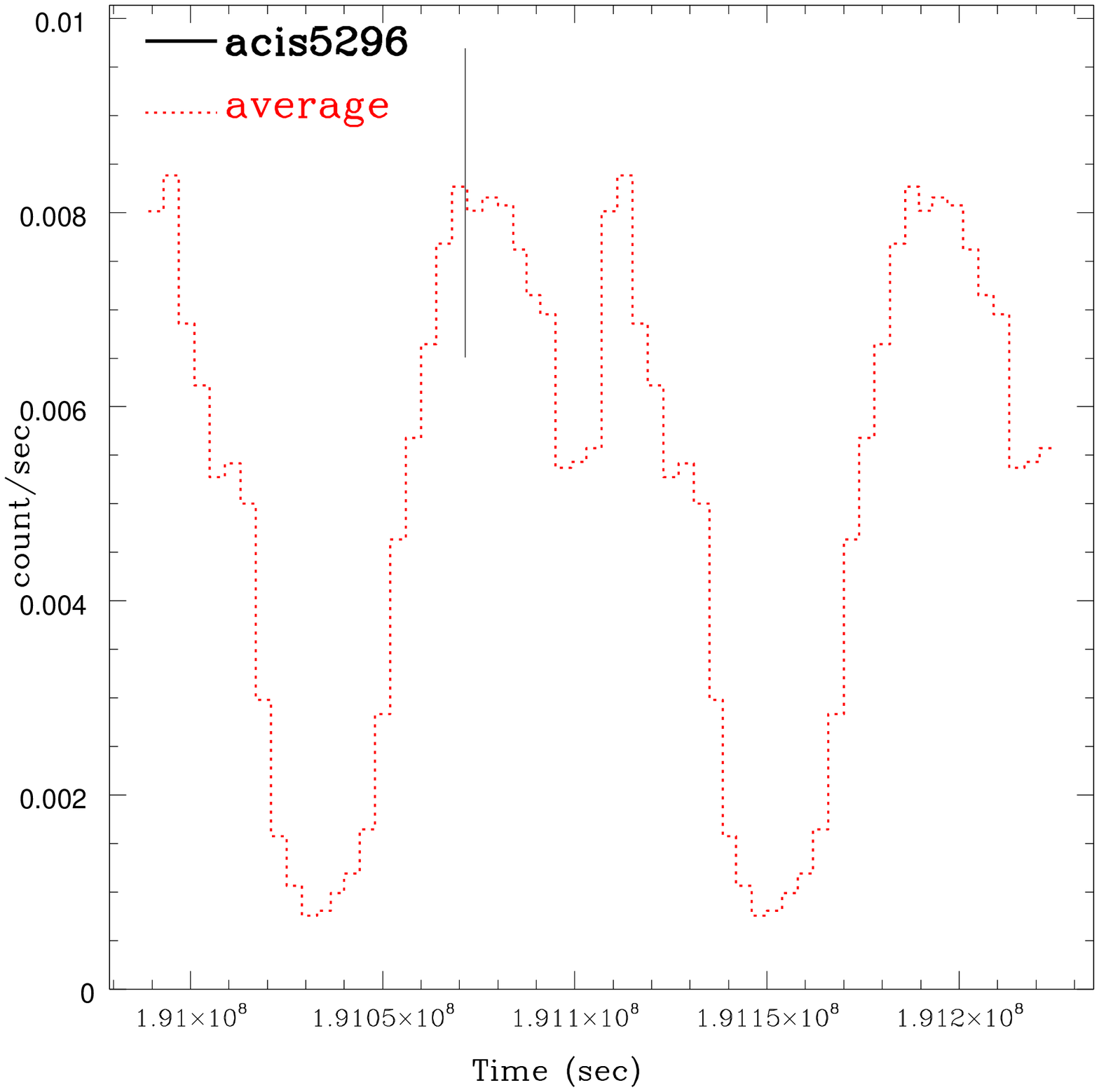}}
{\includegraphics[width=148pt,height=139pt]{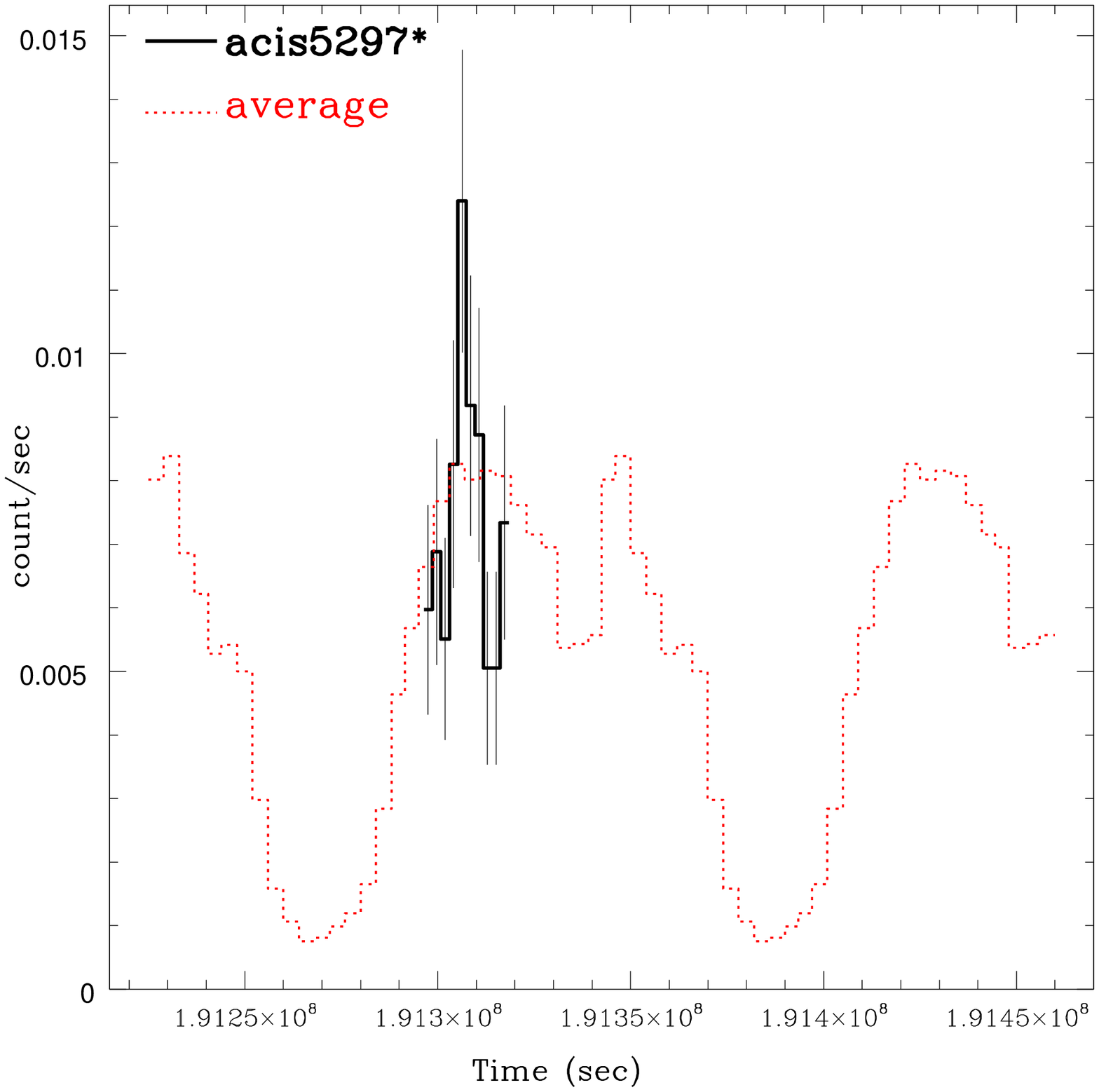}}\\
\vspace{-17pt}

{\includegraphics[width=148pt,height=139pt]{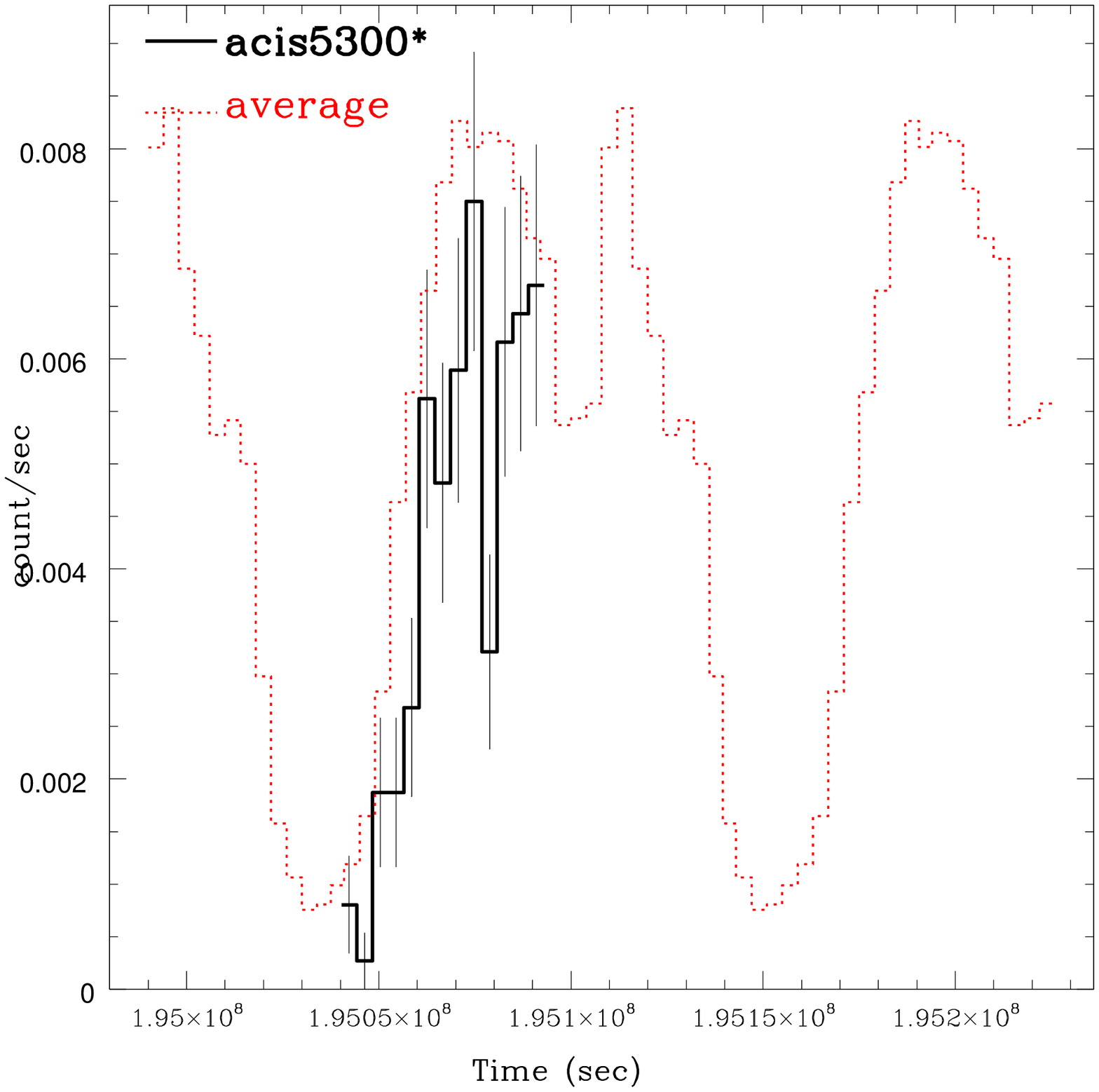}}
{\includegraphics[width=148pt,height=139pt]{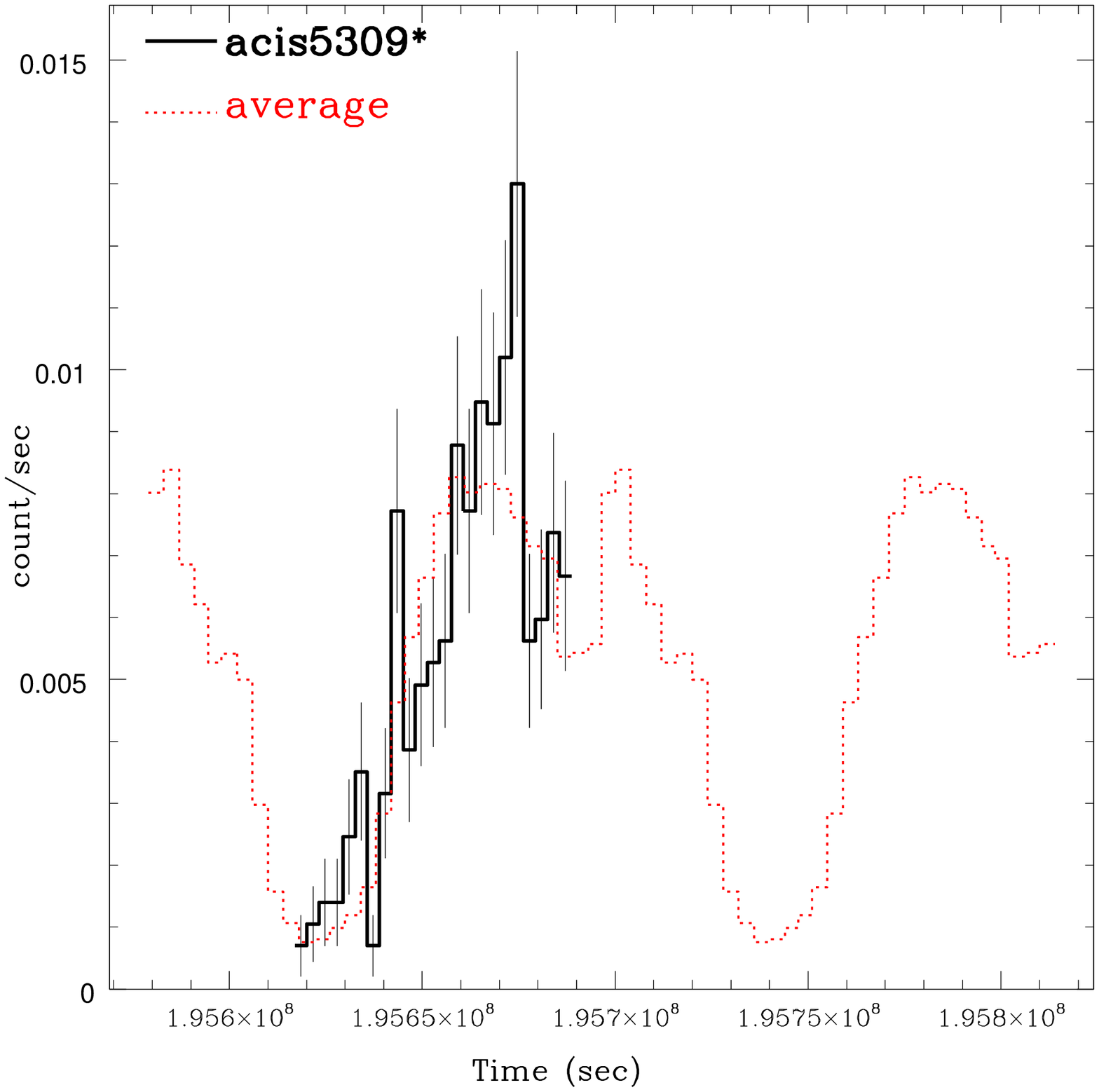}}
{\includegraphics[width=148pt,height=139pt]{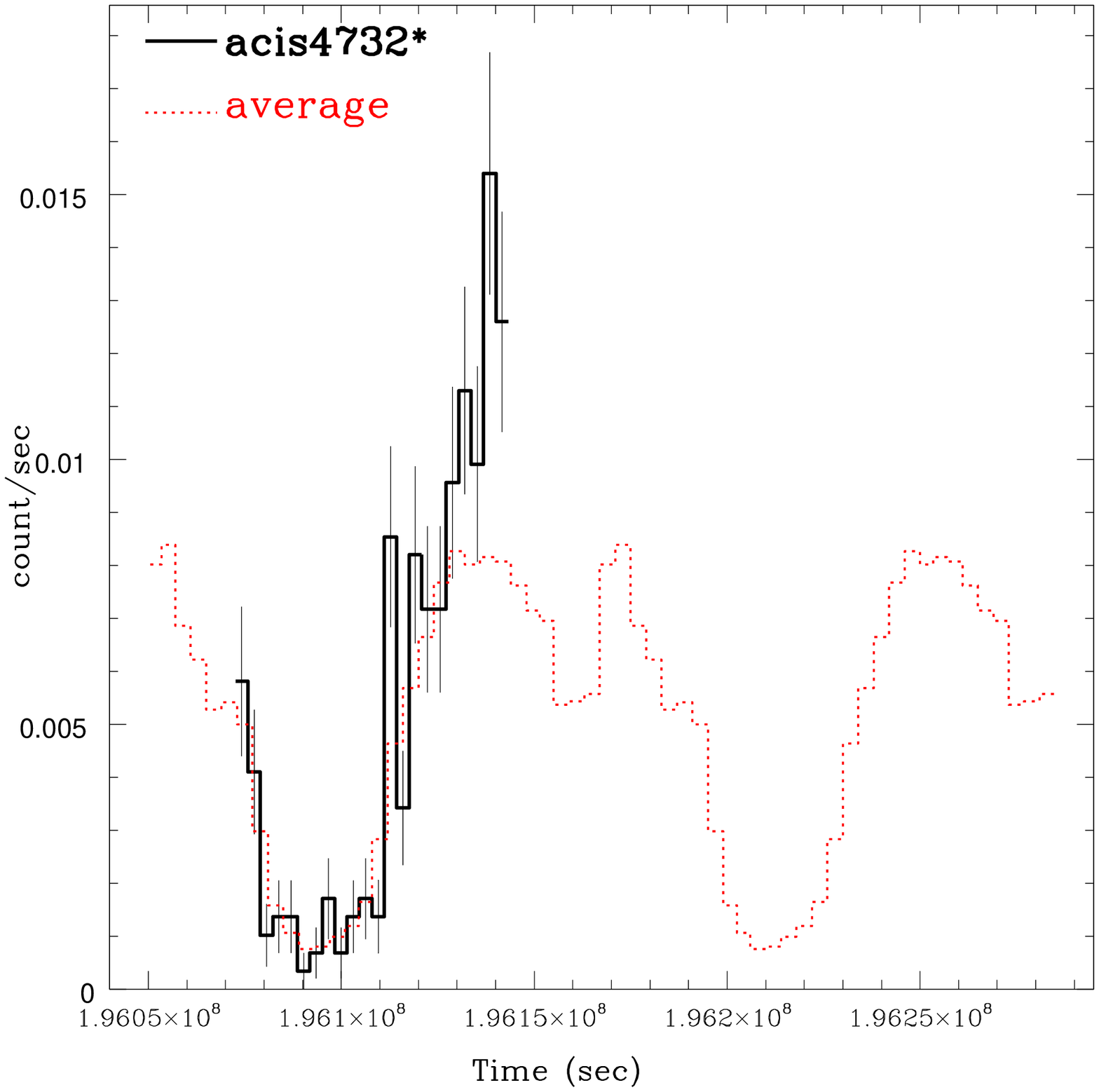}}\\
\vspace{-17pt}

{\includegraphics[width=148pt,height=139pt]{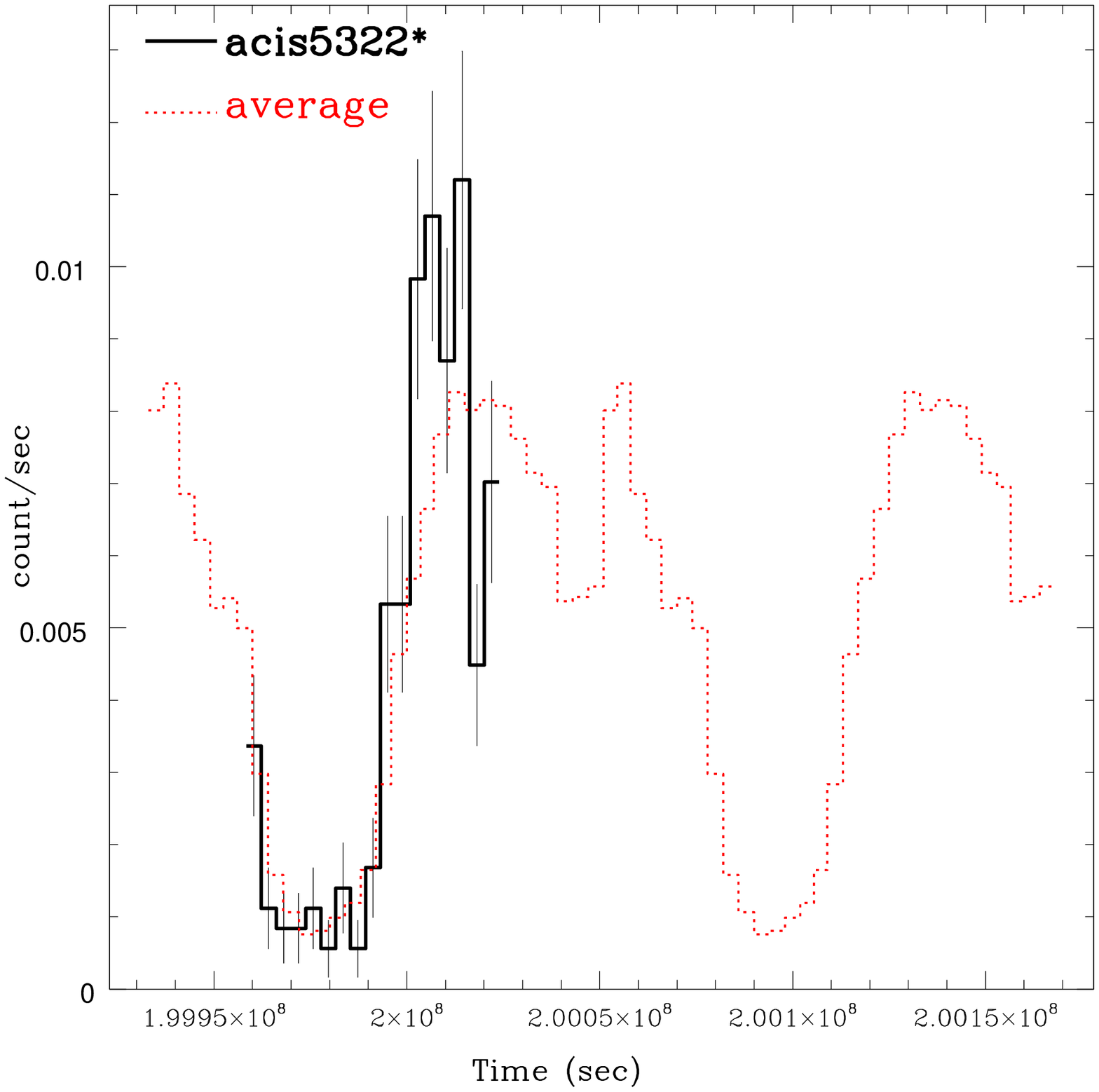}}
{\includegraphics[width=148pt,height=139pt]{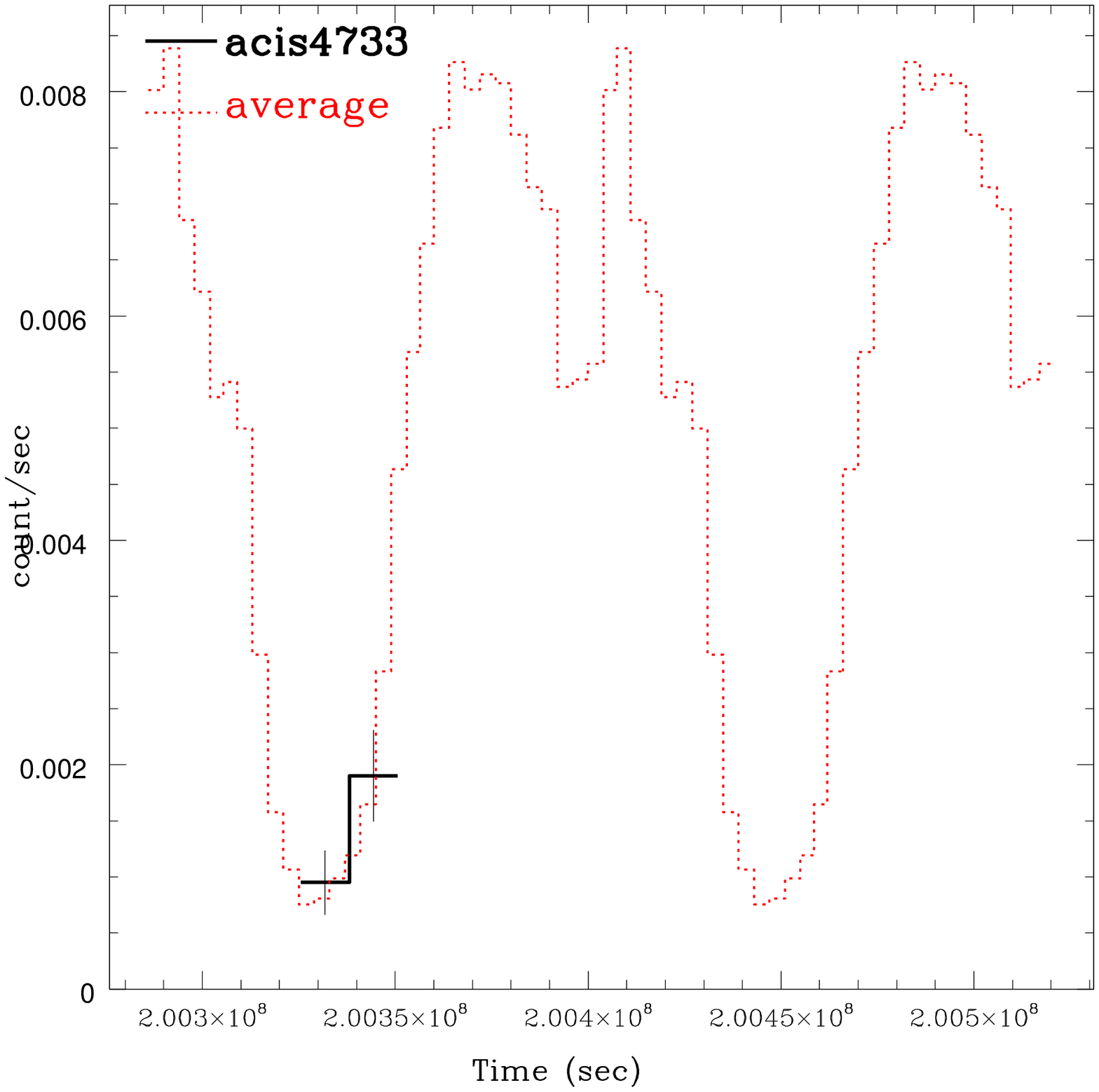}}
{\includegraphics[width=148pt,height=139pt]{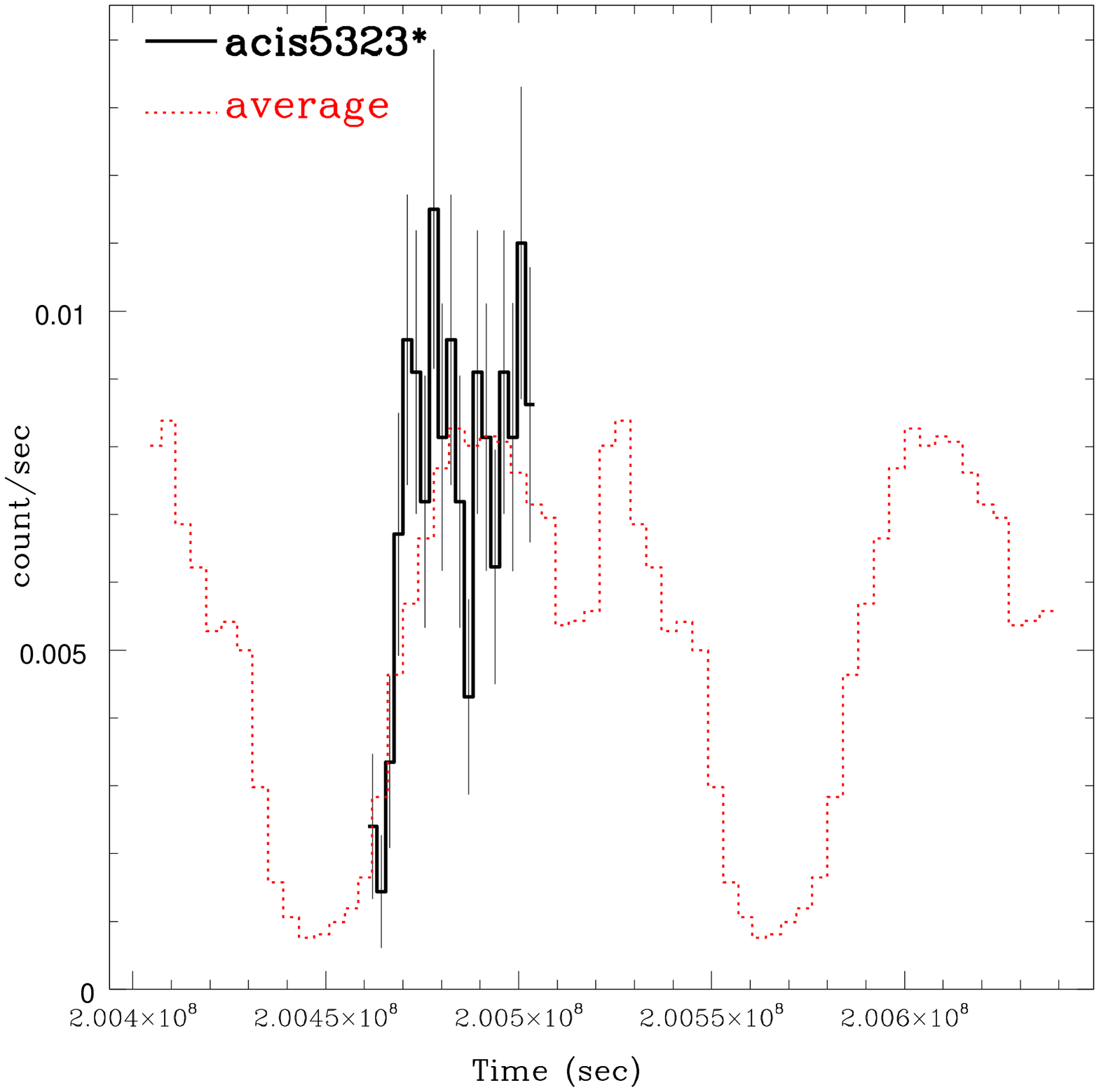}}\\
\vspace{-17pt}

{\includegraphics[width=148pt,height=139pt]{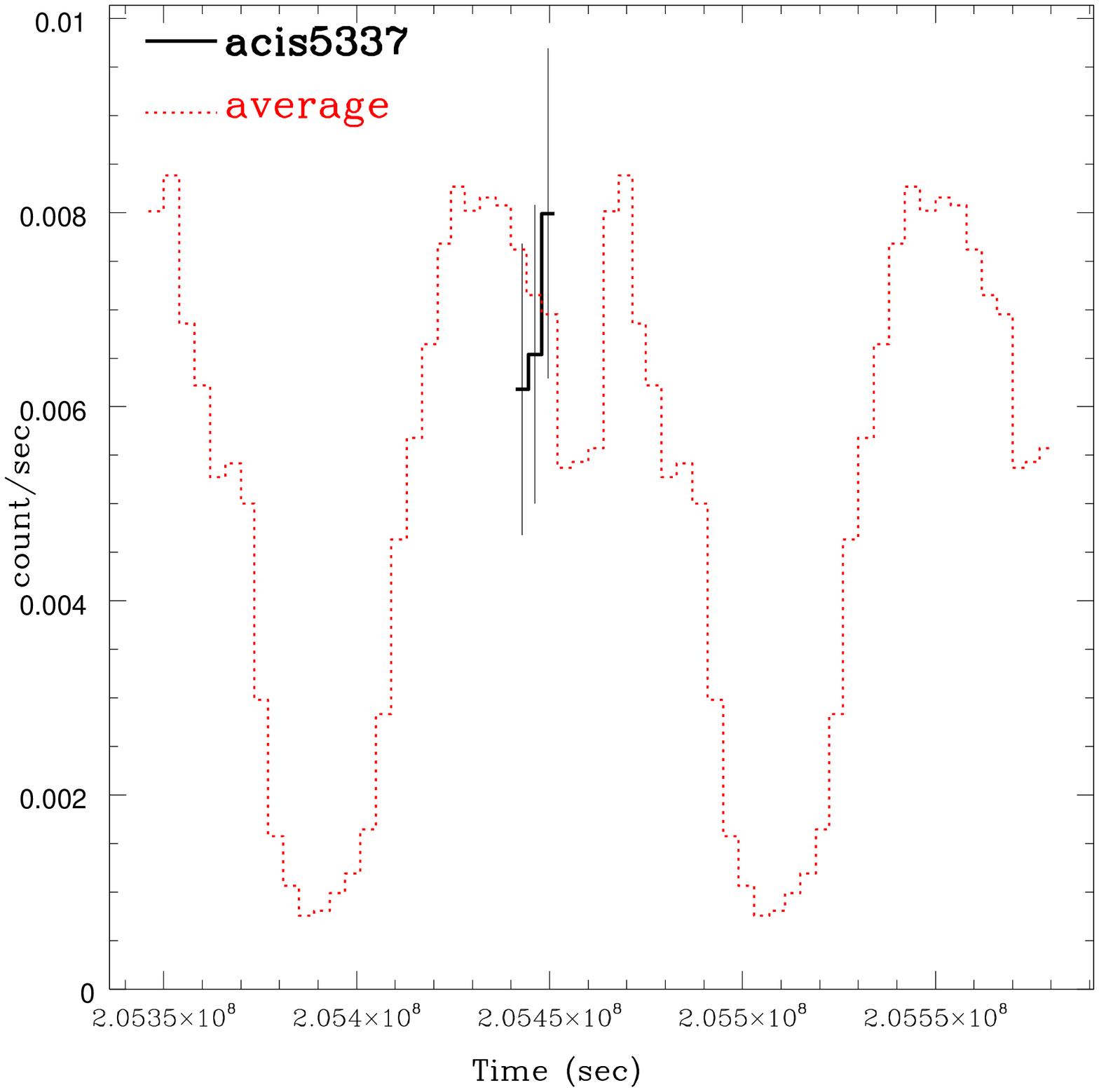}}
{\includegraphics[width=148pt,height=139pt]{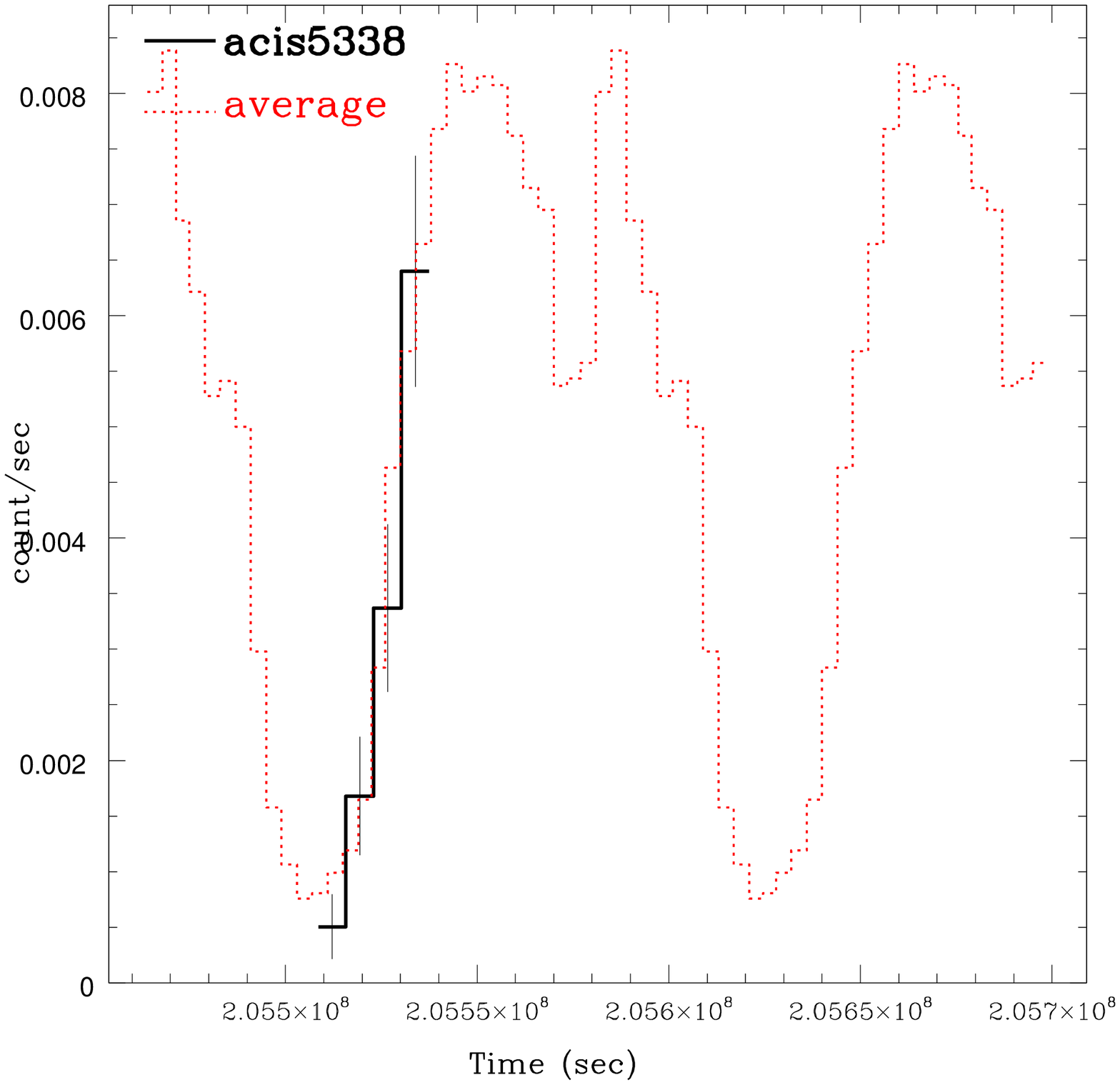}}
{\includegraphics[width=148pt,height=139pt]{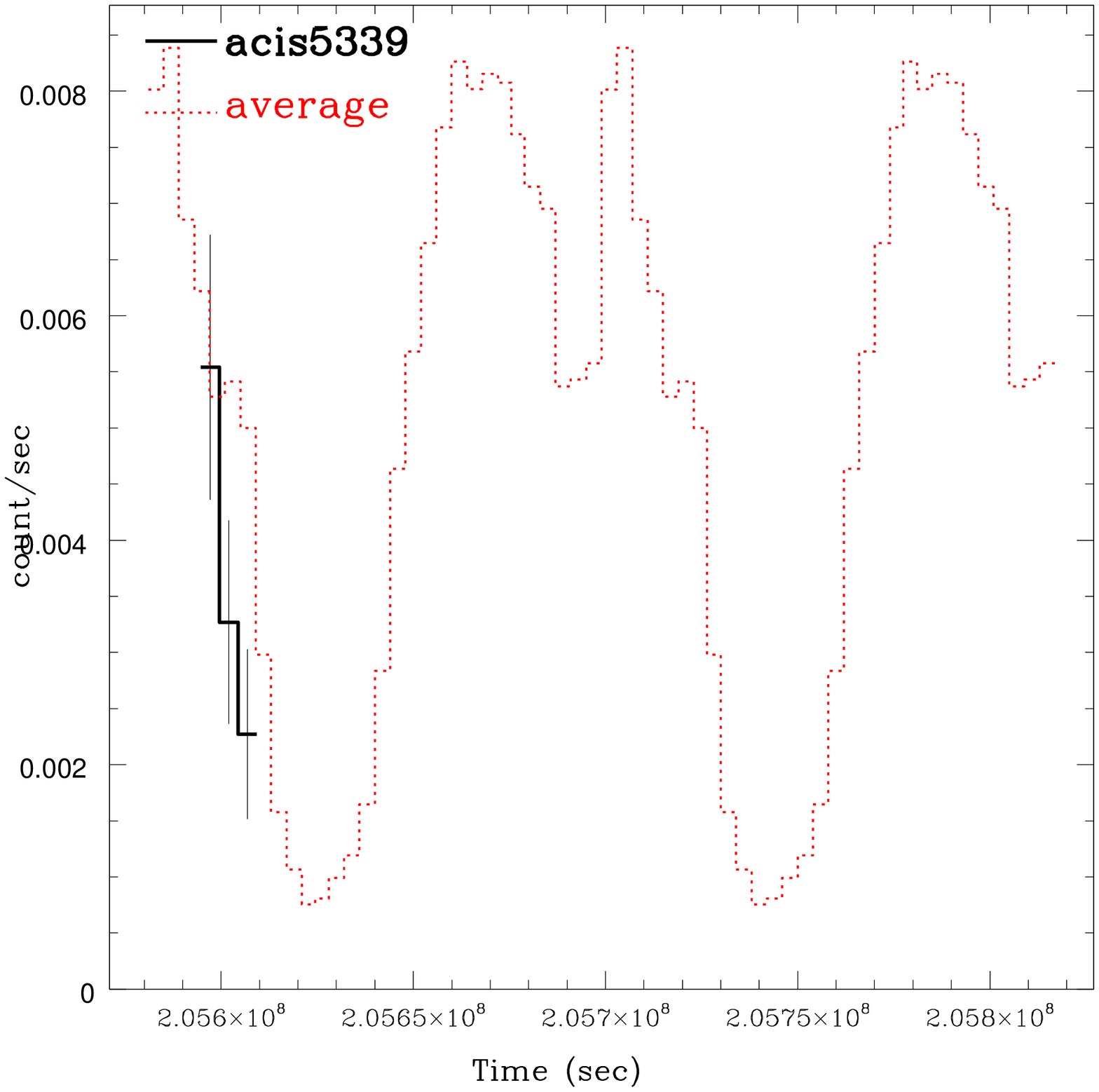}}\\
\vspace{-17pt}
\end{figure}

\begin{figure} 
{\includegraphics[width=148pt,height=139pt]{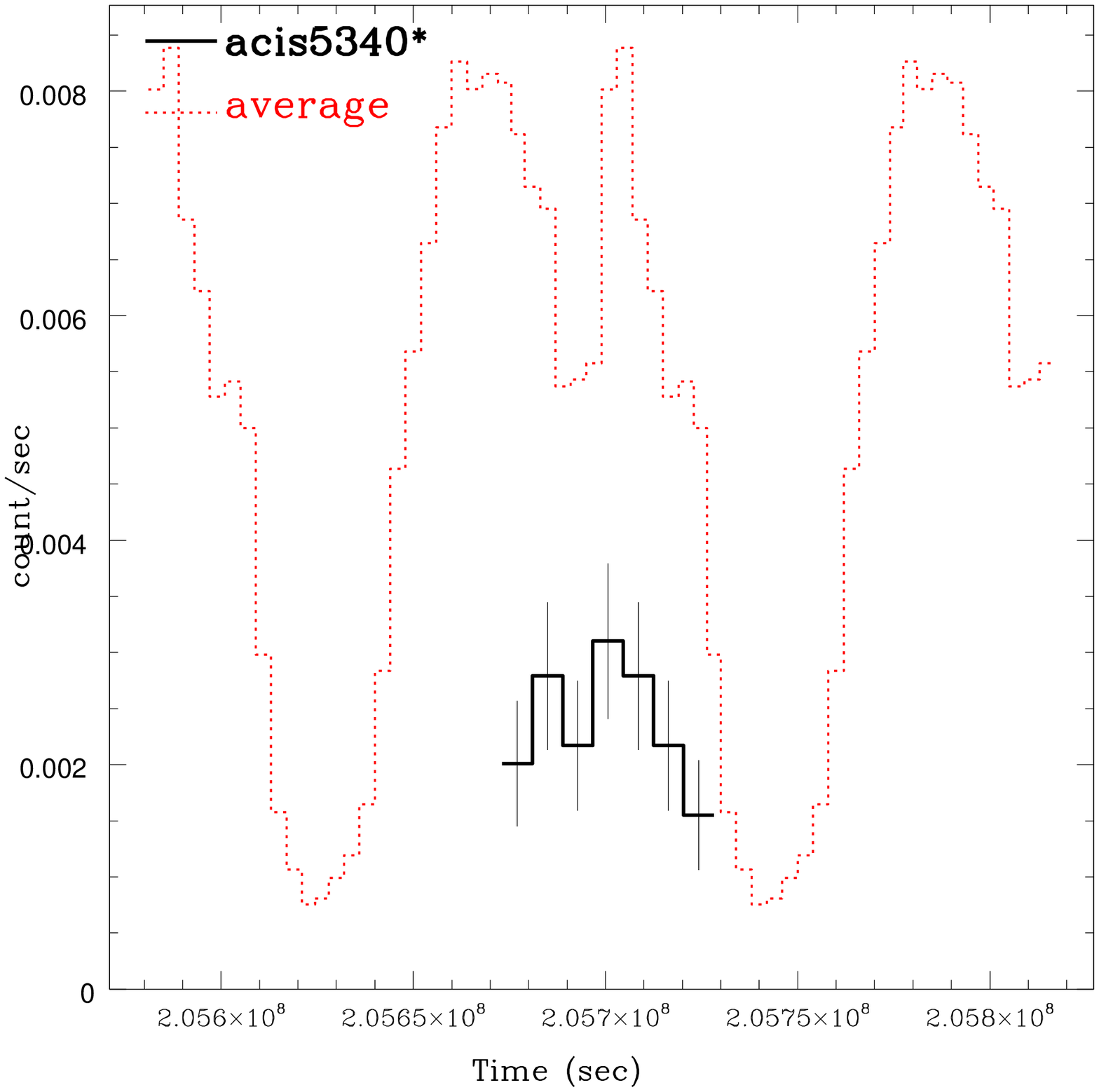}}
{\includegraphics[width=148pt,height=139pt]{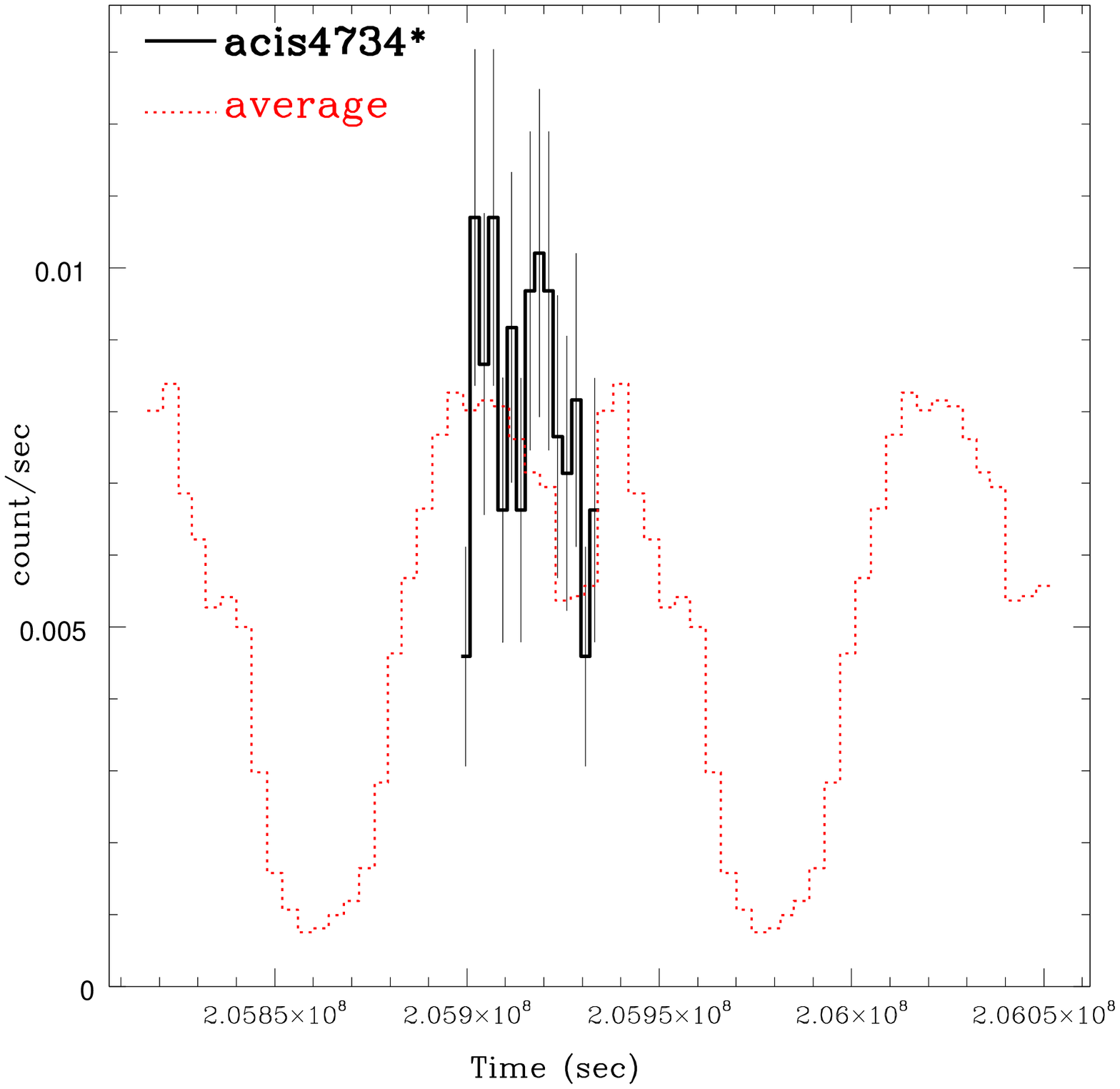}}
{\includegraphics[width=148pt,height=139pt]{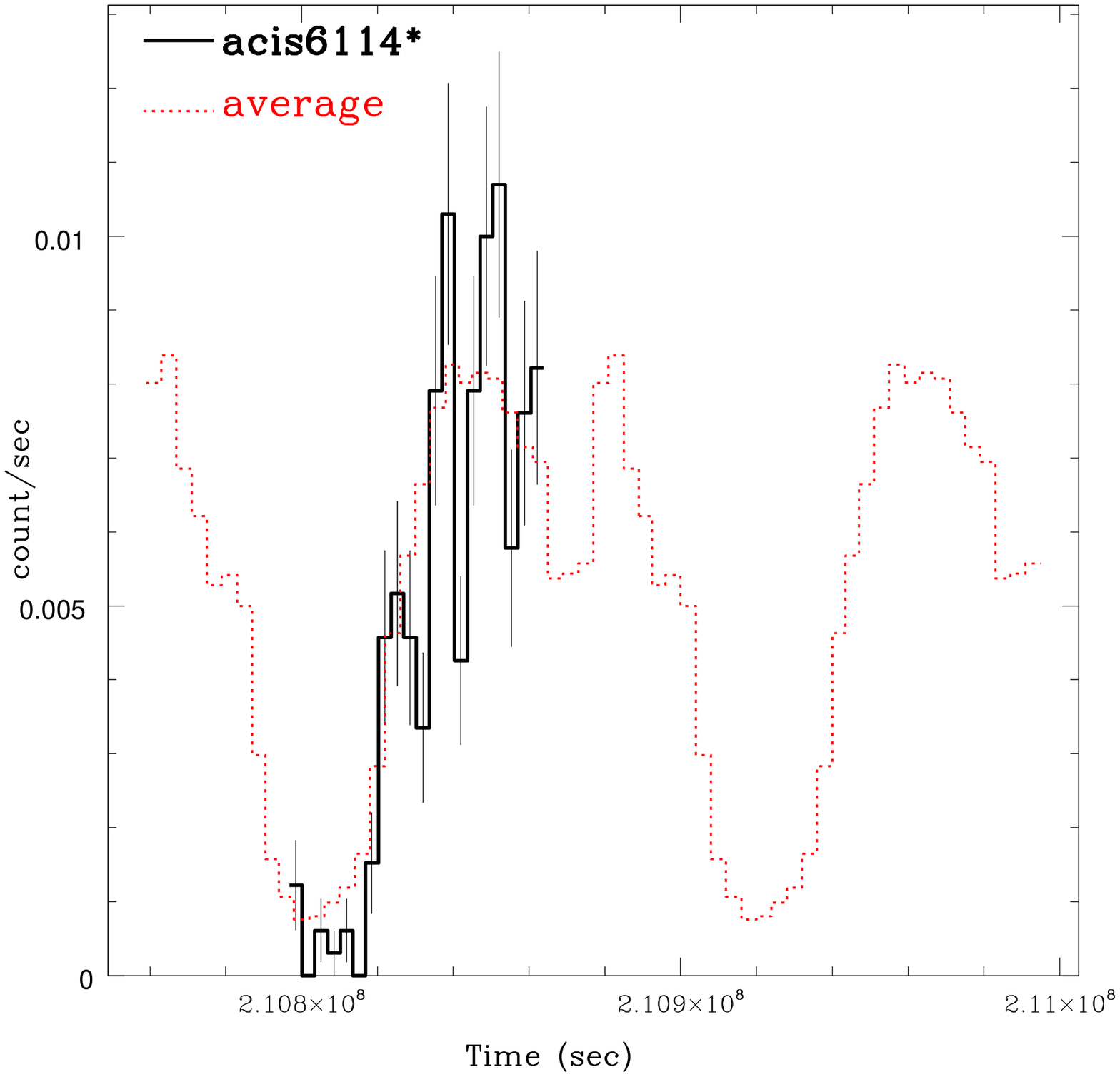}}\\
\vspace{-17pt}

{\includegraphics[width=148pt,height=139pt]{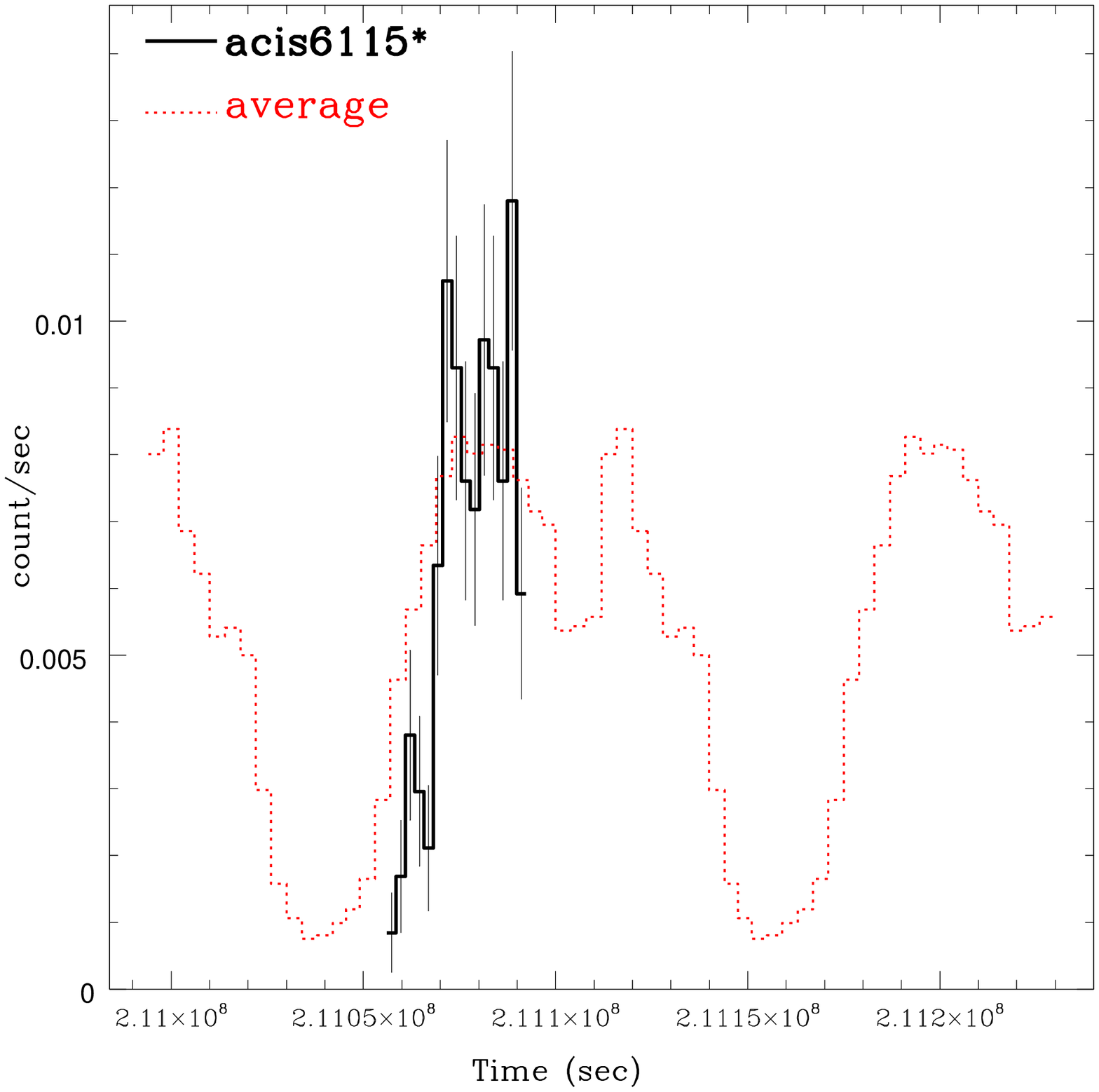}}
{\includegraphics[width=148pt,height=139pt]{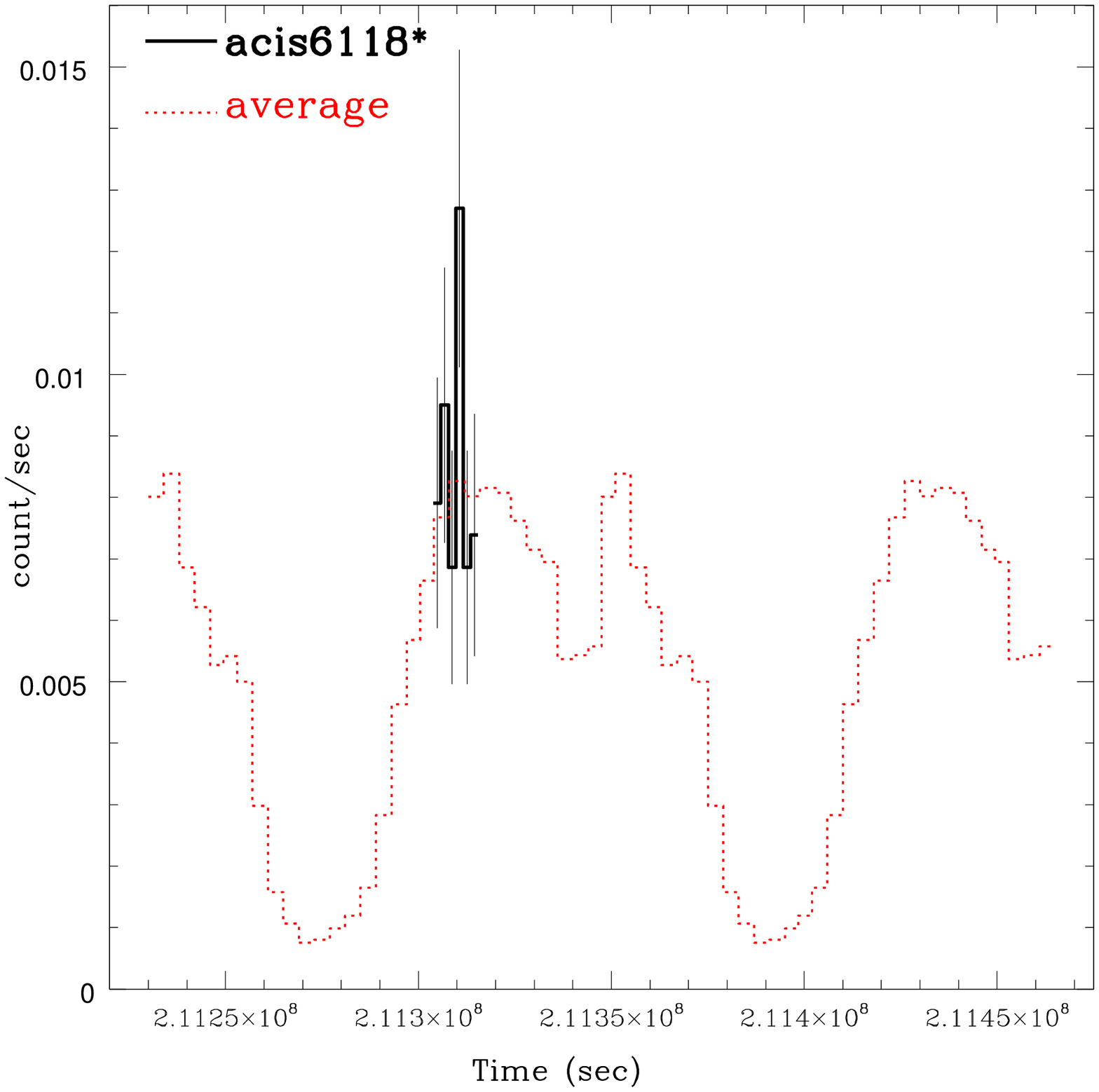}}
{\includegraphics[width=148pt,height=139pt]{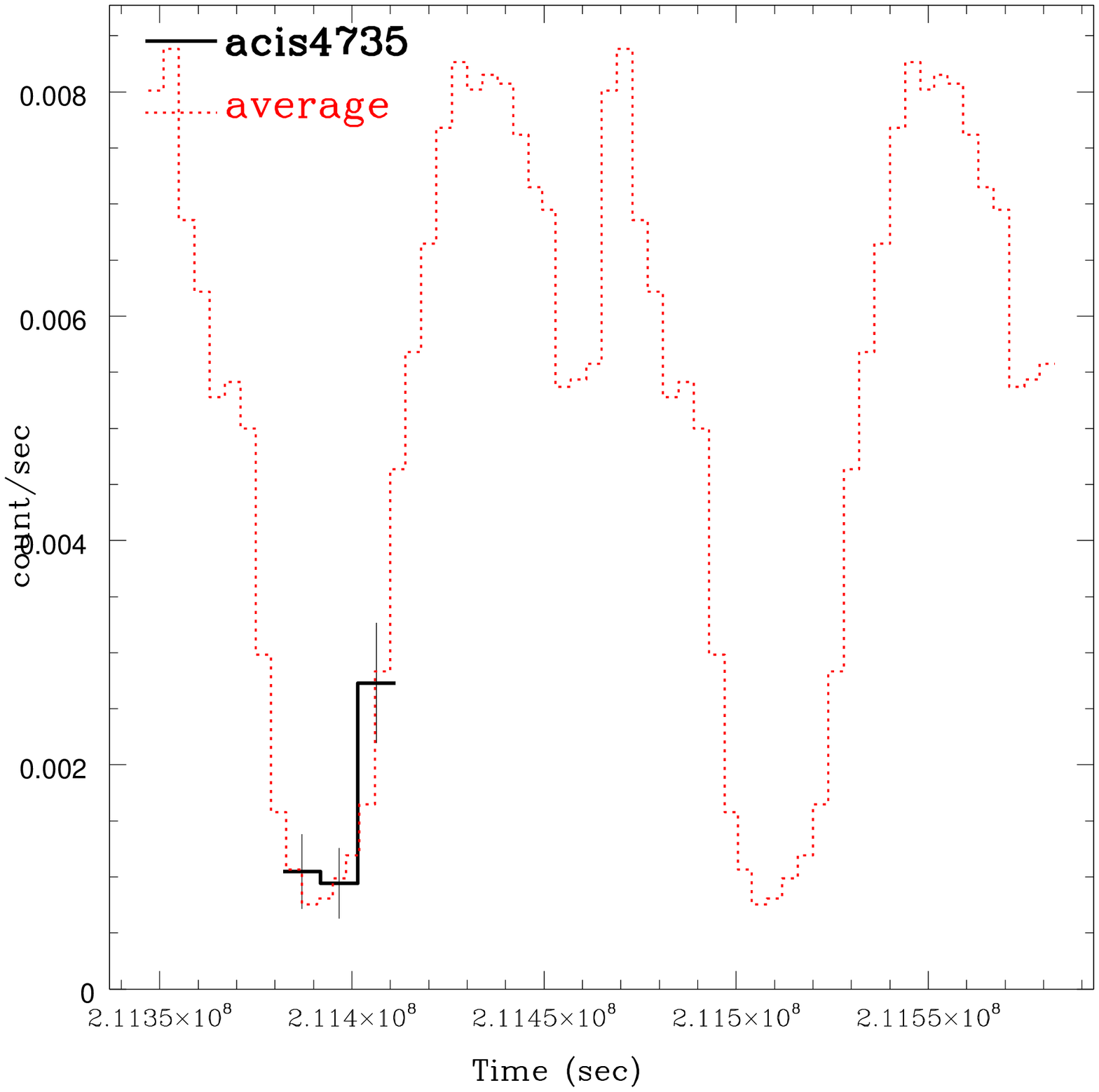}}\\
\vspace{-17pt}

{\includegraphics[width=148pt,height=139pt]{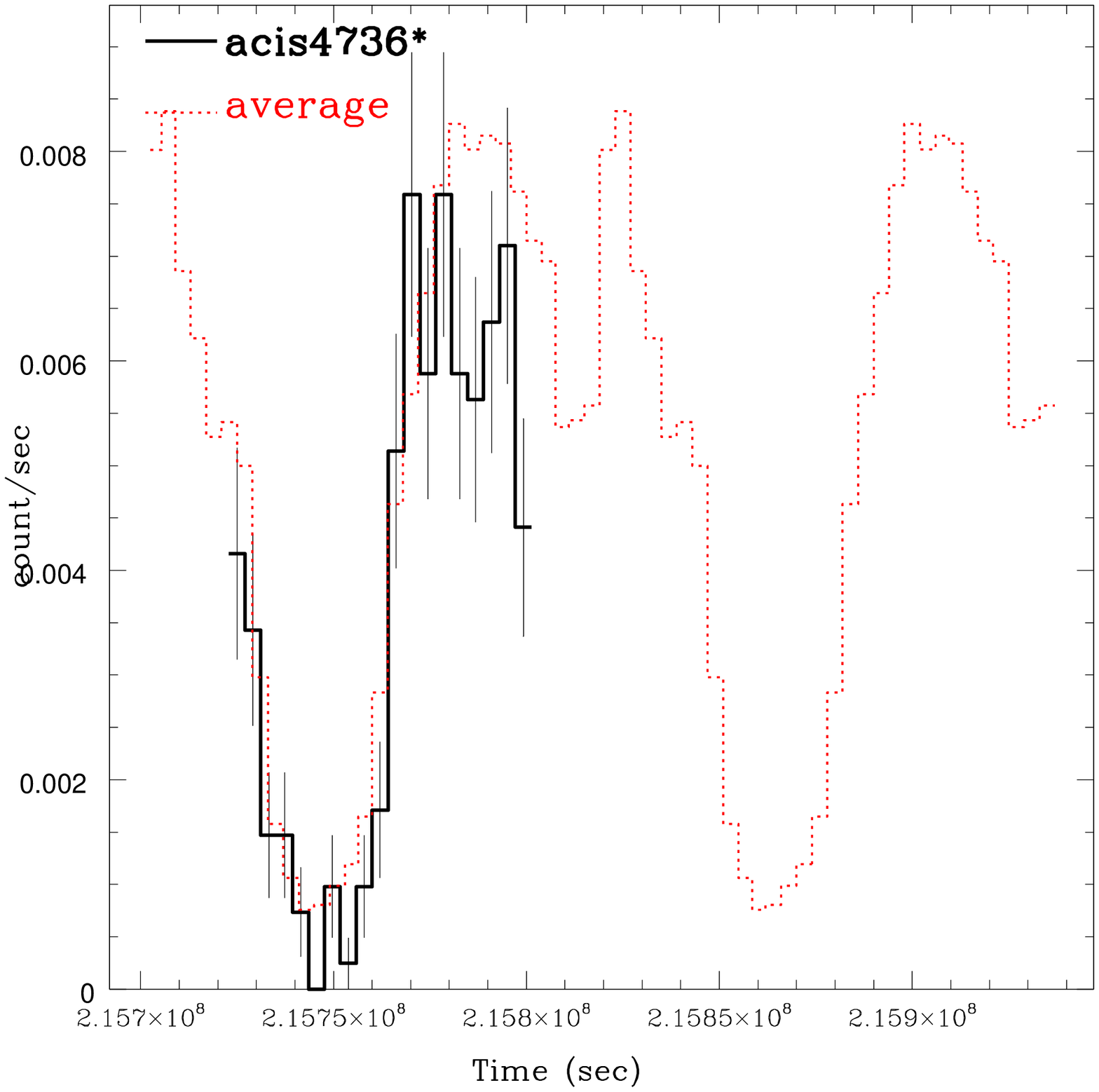}}
{\includegraphics[width=148pt,height=139pt]{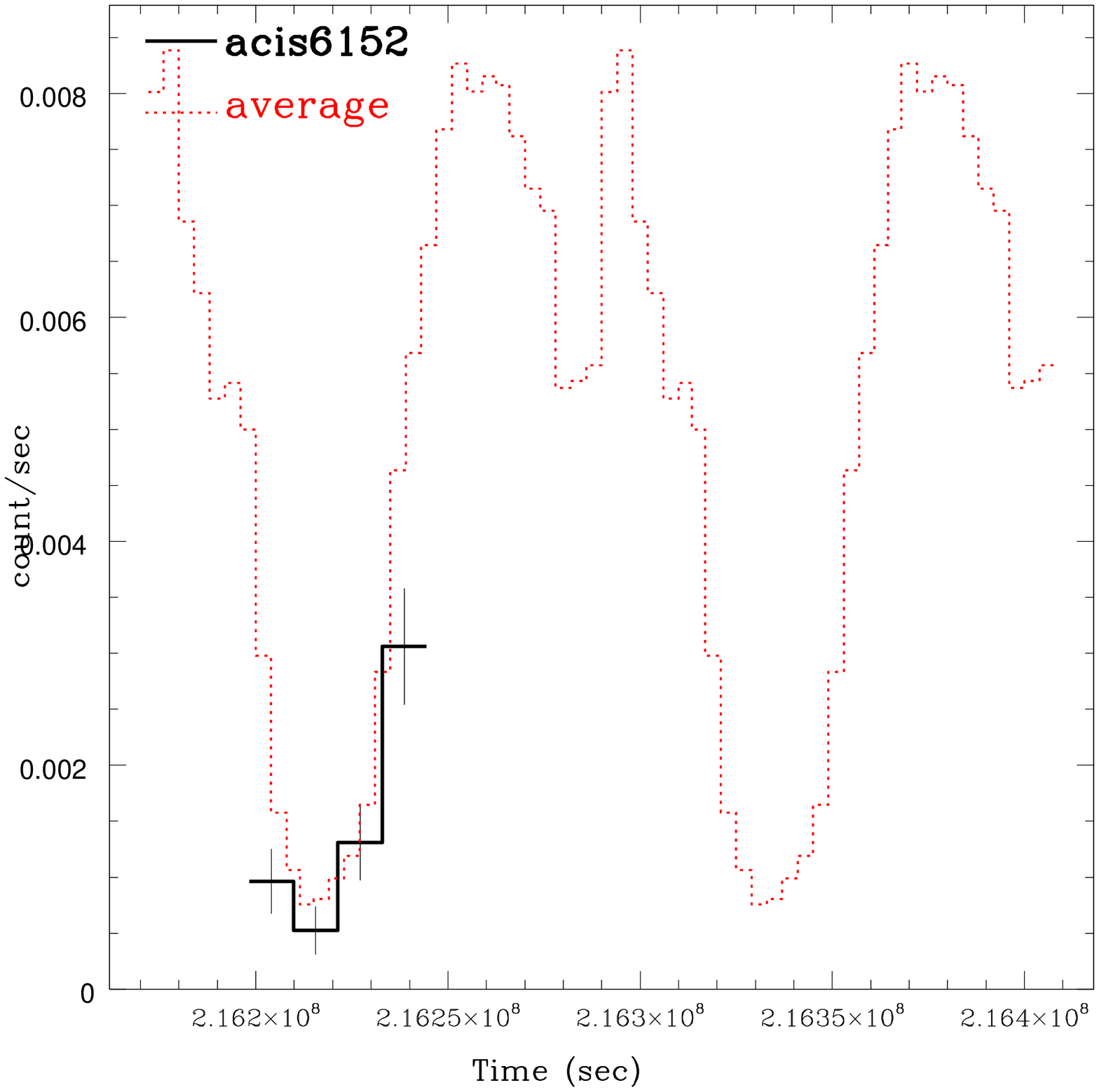}}
{\includegraphics[width=148pt,height=139pt]{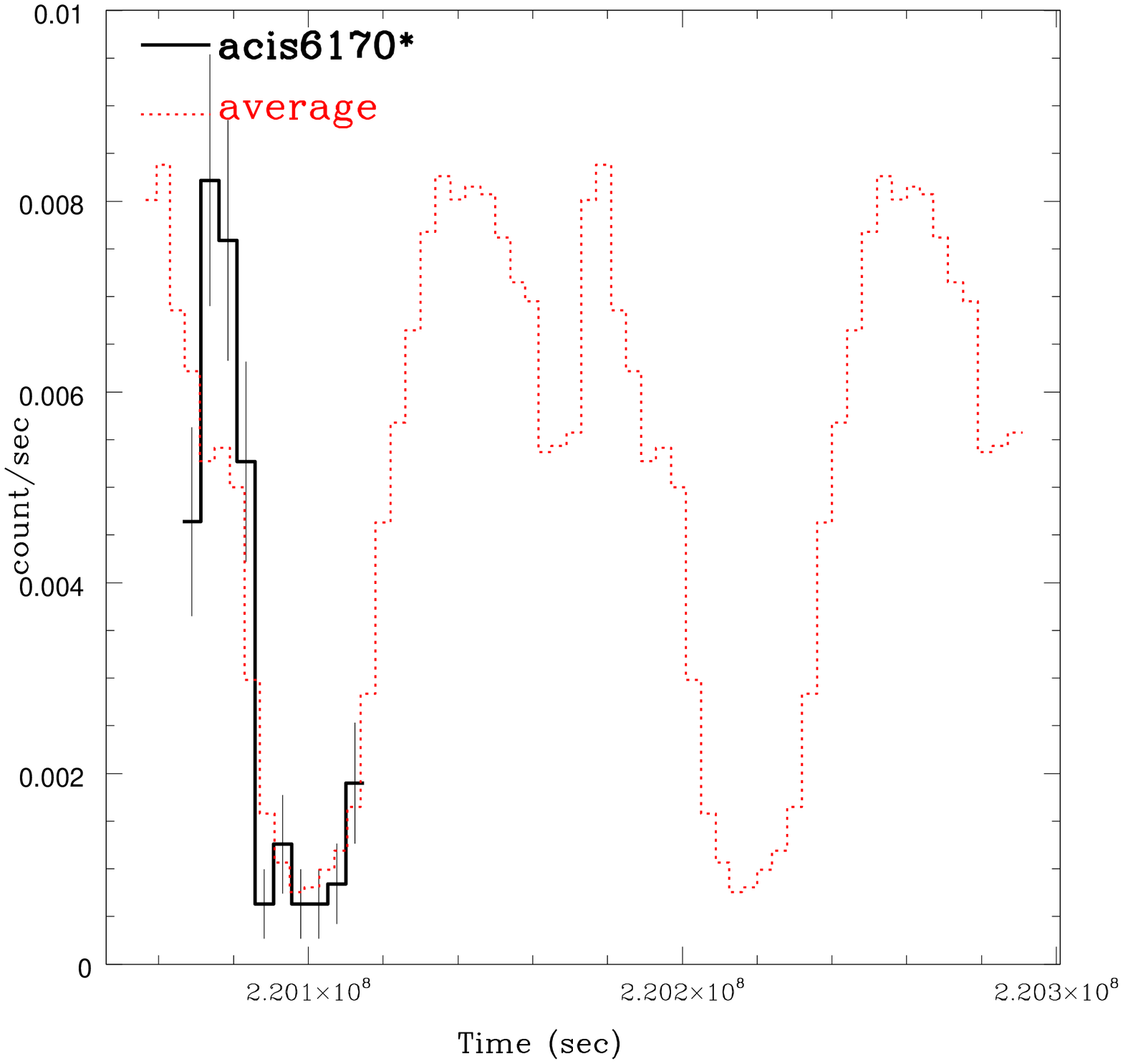}}\\
\vspace{-17pt}

{\includegraphics[width=148pt,height=139pt]{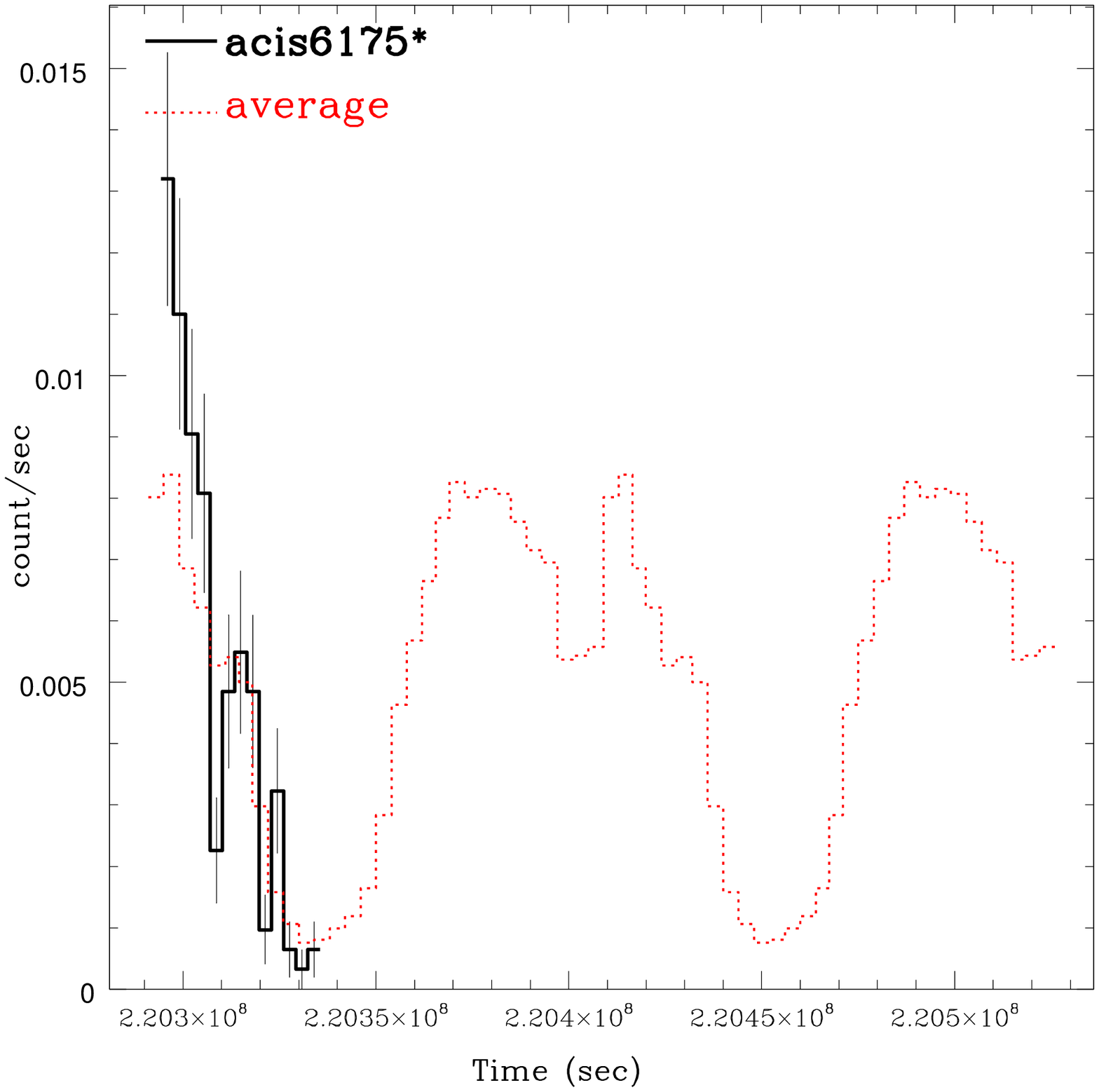}}
{\includegraphics[width=148pt,height=139pt]{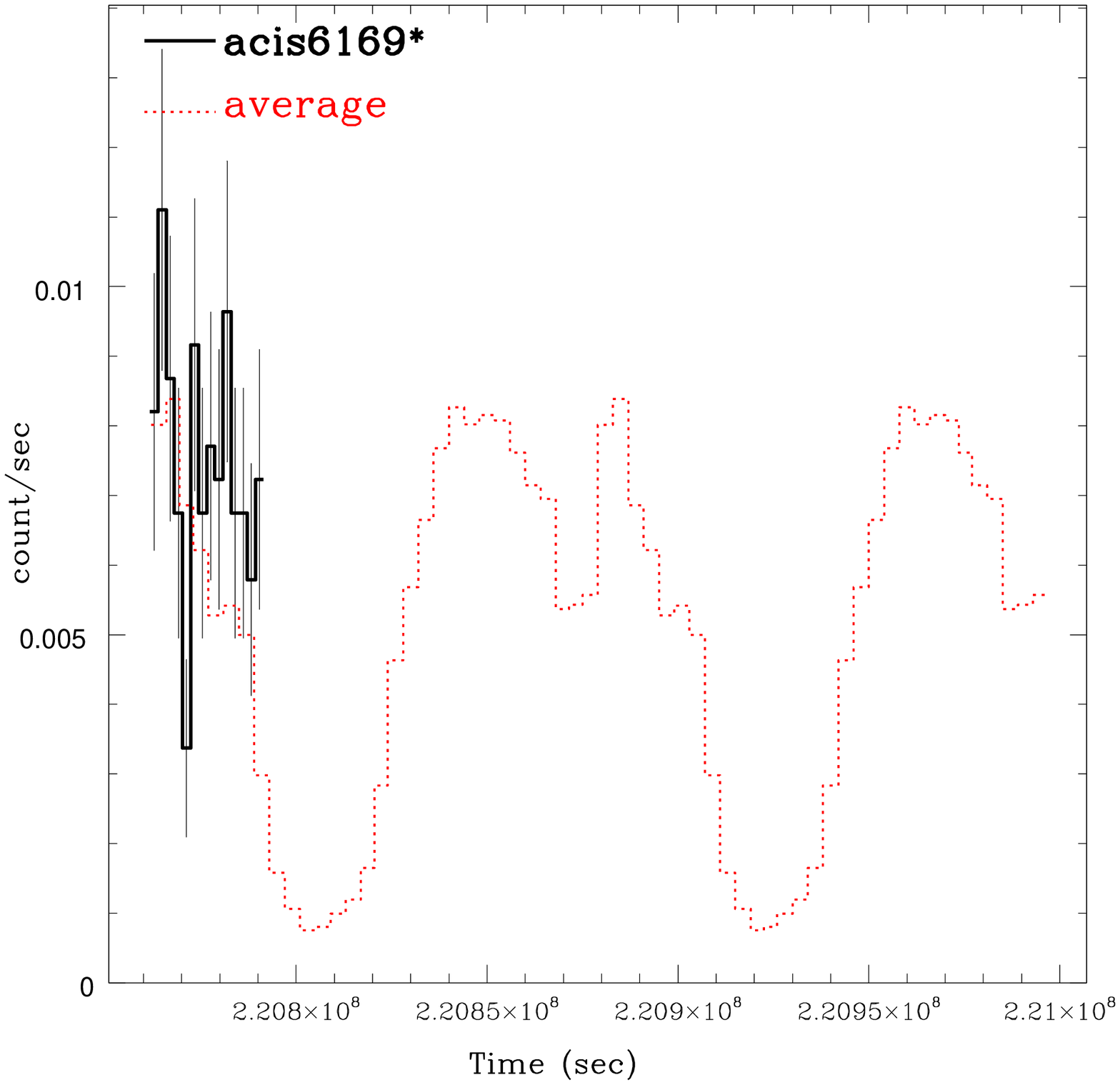}}
{\includegraphics[width=148pt,height=139pt]{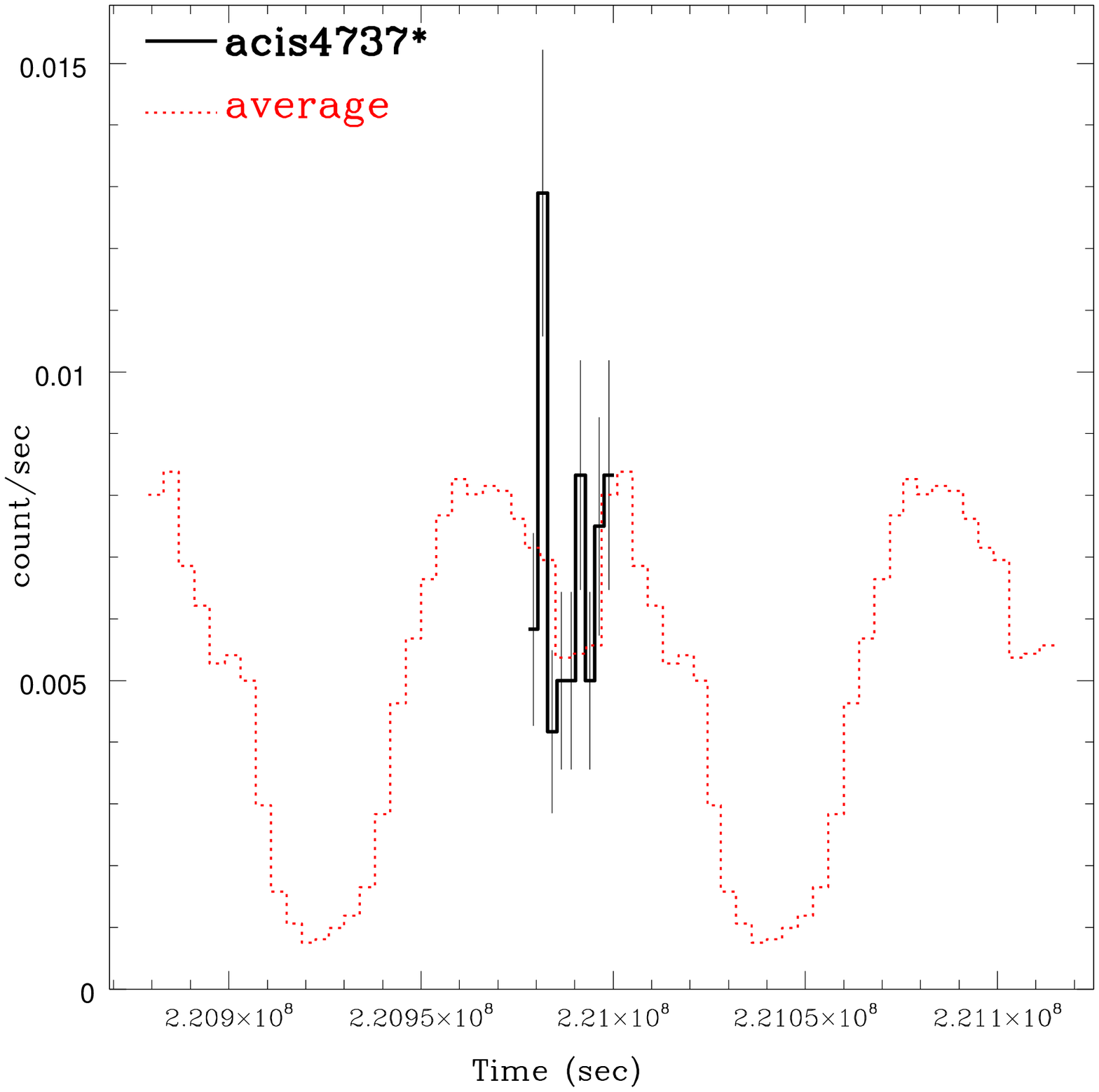}}\\
\vspace{-17pt}

\caption{The individual light curves from the M101 ultra-deep survey in
comparison to the average light curve for the period of 32.69 hours. The light
curves with a ``*'' suffixed in the key are the 18 ``good'' light curves used
to search for the period. }

\end{figure}


\begin{figure}
\plotone{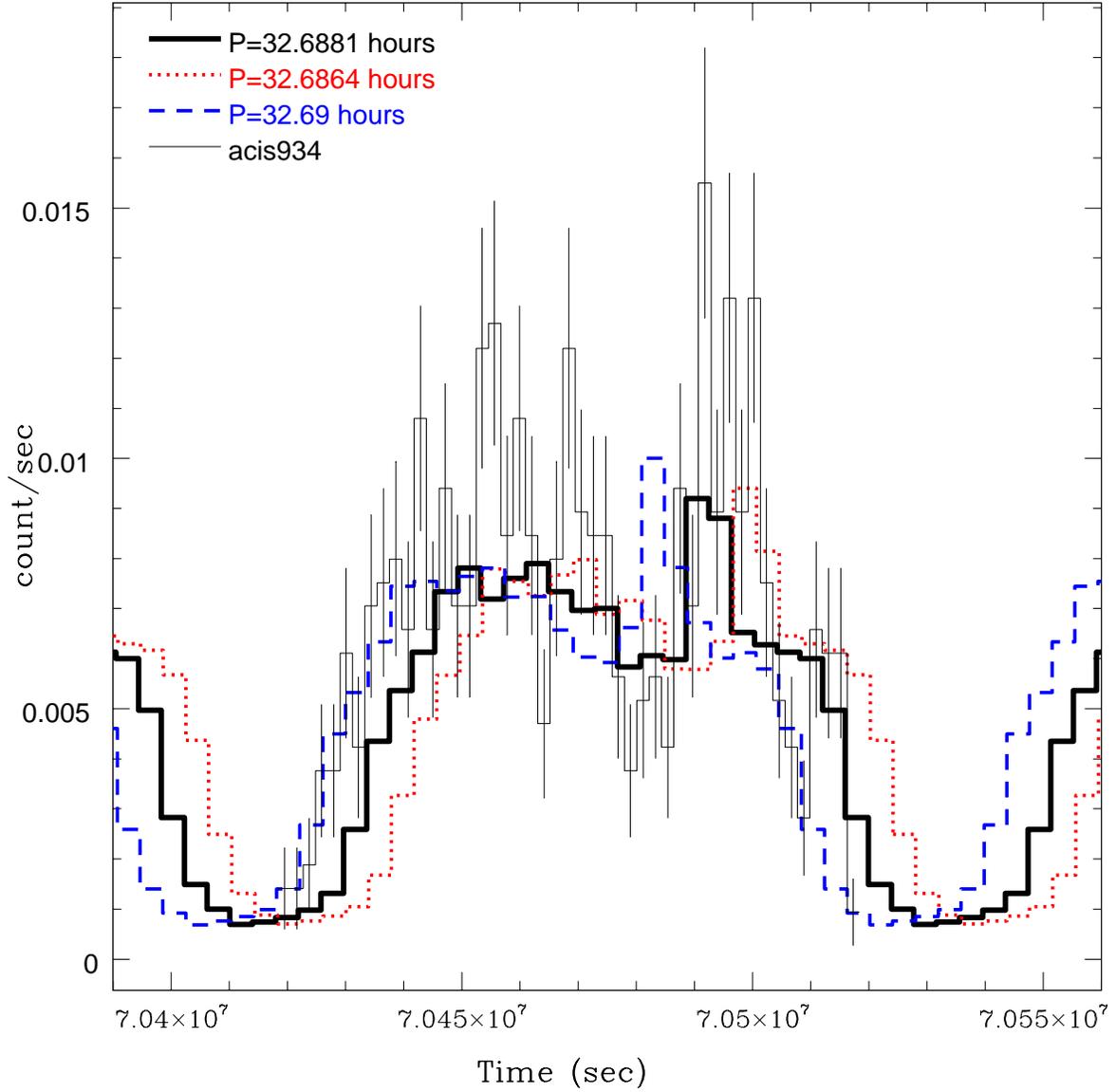}
\caption{The light curve for ObsID 934 in comparison with the average light
curves for slightly different periods. The light curve are mostly consistent
with the average light curve for the period of 32.688 hours (thick solid) by
aligning the eclipse ingress and the minor dip.  }

\end{figure}


\begin{figure}
\plotone{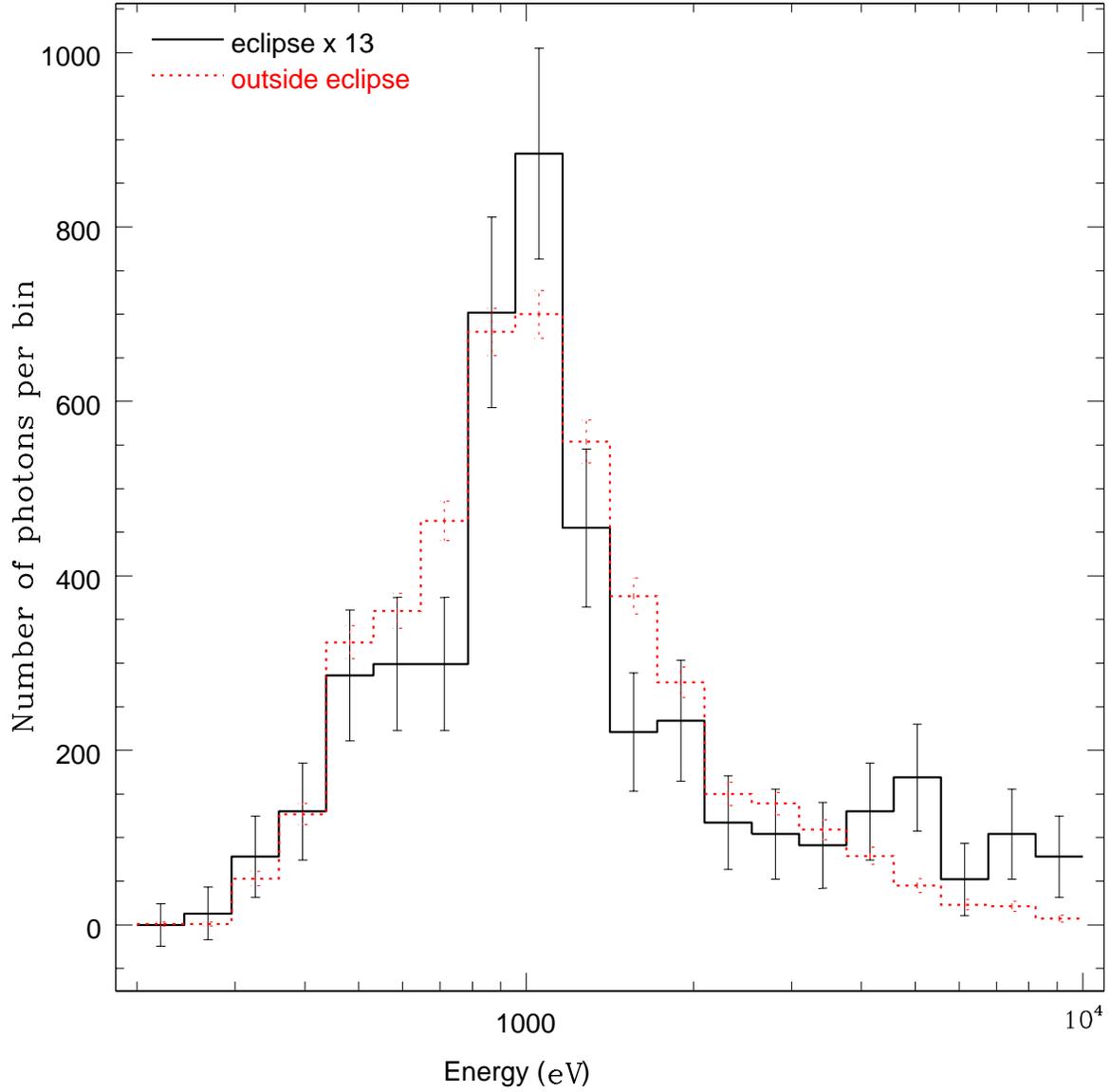}
\caption{The energy distributions for photons within the eclipses and outside
the eclipses from all 25 observations. The coadded spectrum within the eclipses
exhibited a lack of soft photons below 2 keV and an excess of hard photons
above 4 keV as compared to the coadded spectrum outside the eclipses. }

\end{figure}

\begin{figure}
\plotone{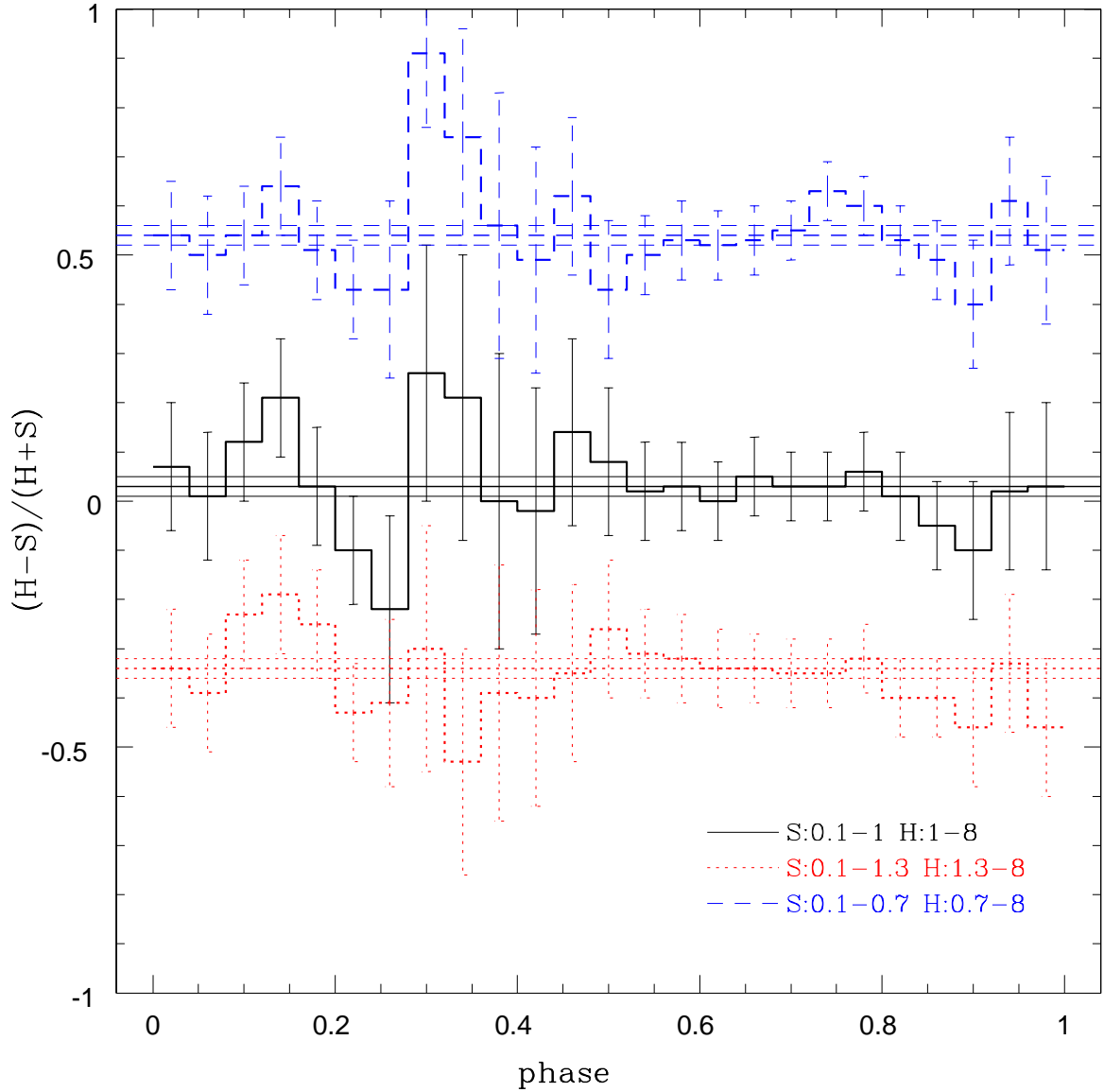}
\caption{The hardness ratios as a function of phases. The hardness ratio was
calculated as (H-S)/(H+S). Phase bins of 0.04 were used to ensure at least 40
photons per bin. The straight lines are the average hardness ratios for three
band sets and their errors.}

\end{figure}

\begin{figure}
\plottwo{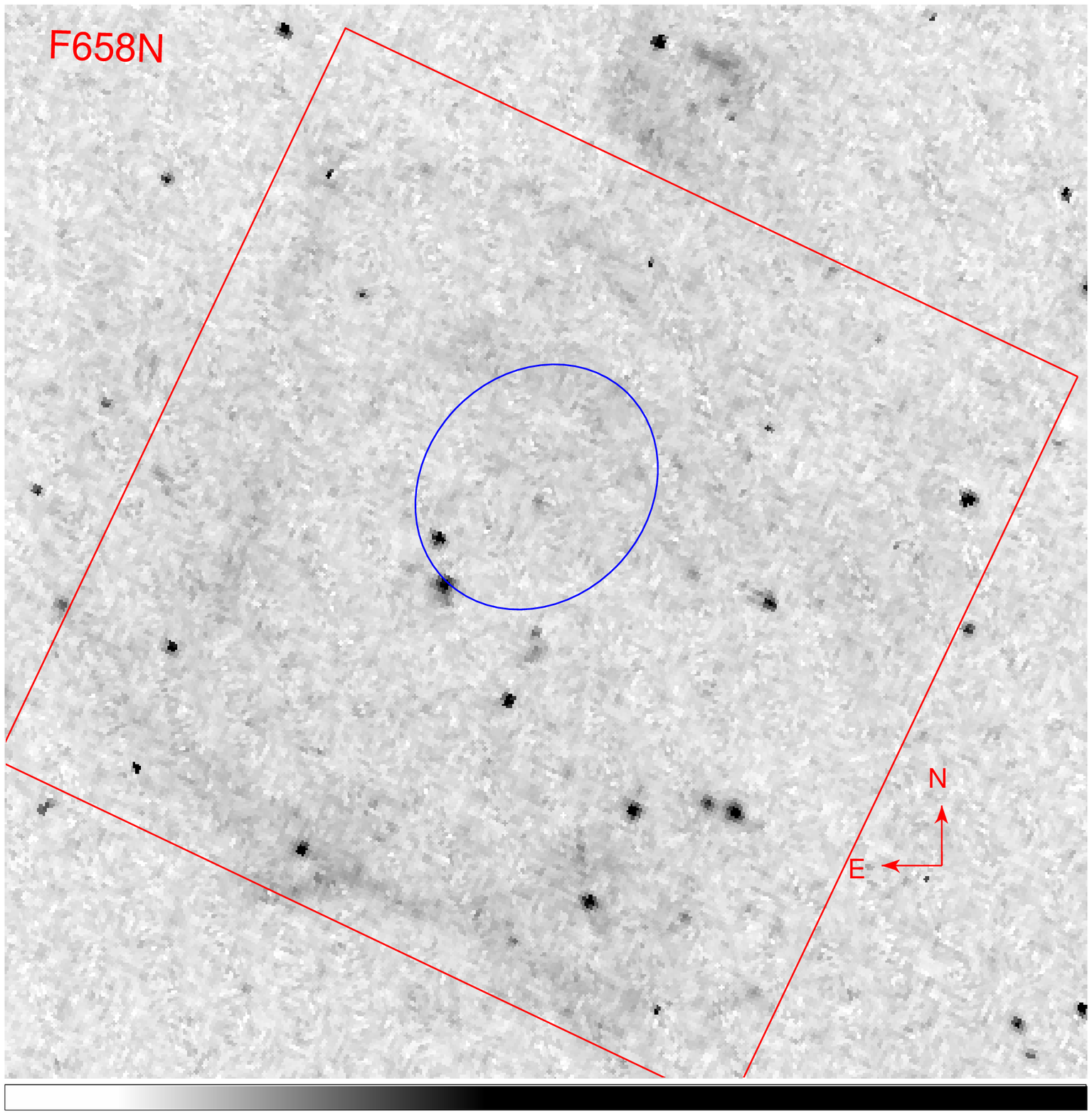}{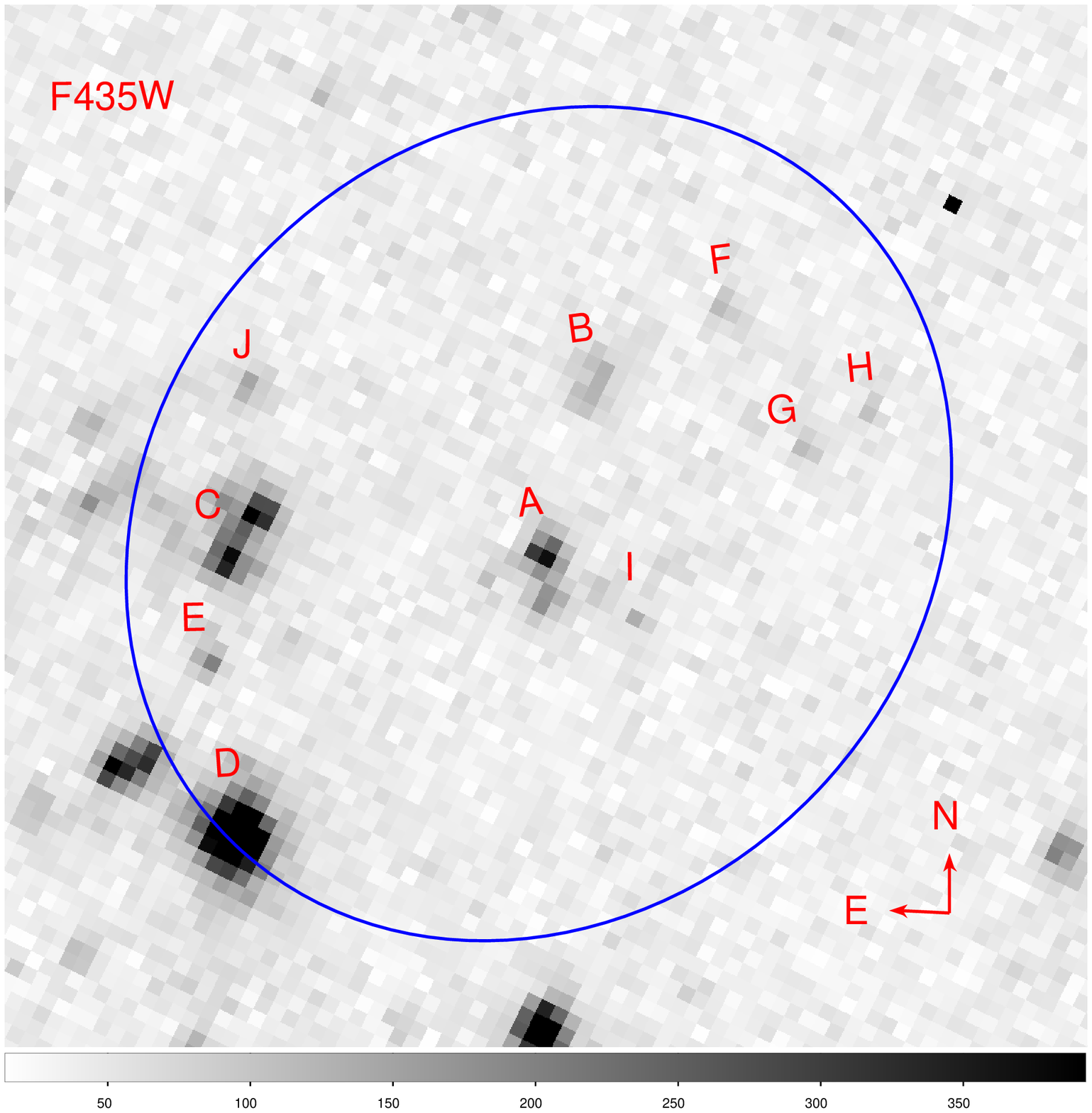}
\end{figure}
\begin{figure}
\plottwo{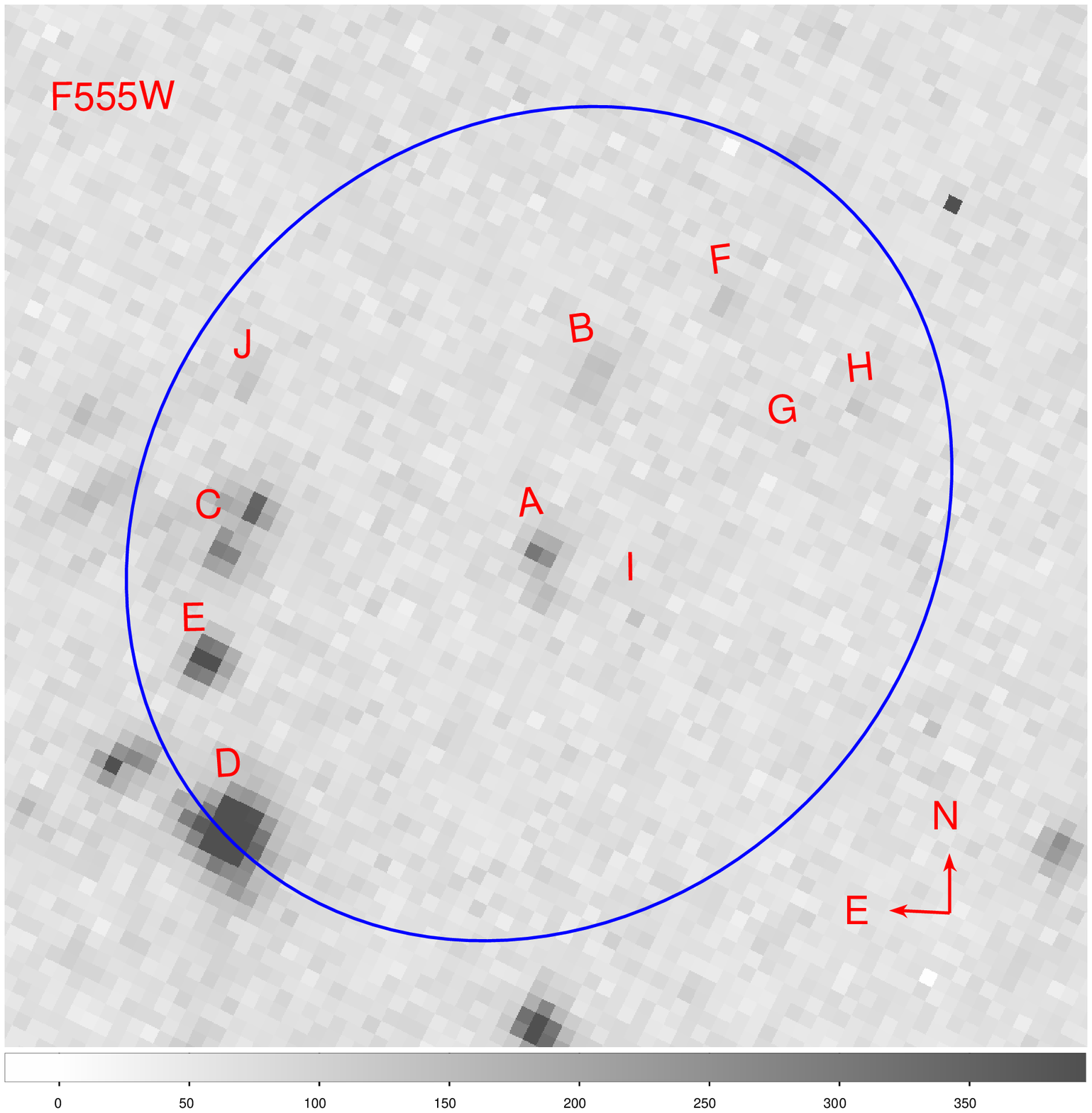}{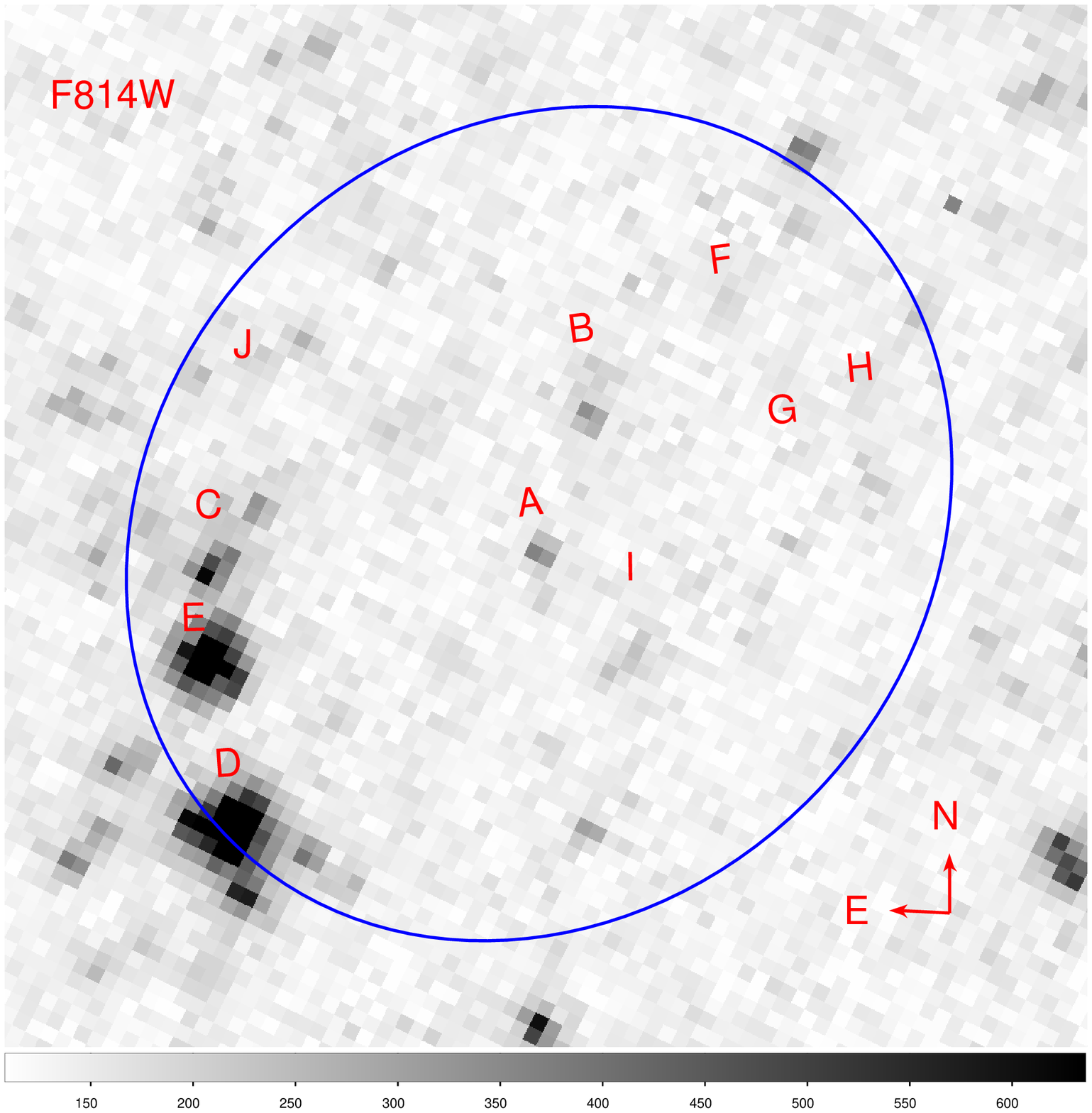}
\caption{(a) The HST ACS/WFC $H_\alpha$ image for X7. The $9^{\prime\prime}
\times 9^{\prime\prime}$ box encloses the irregular shell structure of the
super-bubble. The ellipse centered at the corrected X-ray position of X7
encloses 95\% of the source photons as reported by WAVDETECT.  (b) The HST
ACS/WFC B image for X7. The ellipse is the same as in the $H_\alpha$ image. The
letters label the concentrations of bright stars. (c) The HST ACS/WFC V image
for X7. (d) The HST ACS/WFC I image for X7. The overplotted ellipse has a size
of $1\farcs4 \times 1\farcs2$, while the corrected X-ray position of X7 has an
error of about $0\farcs3$ that only encloses stars A1-A3. }

\end{figure}


\begin{figure}
\plotone{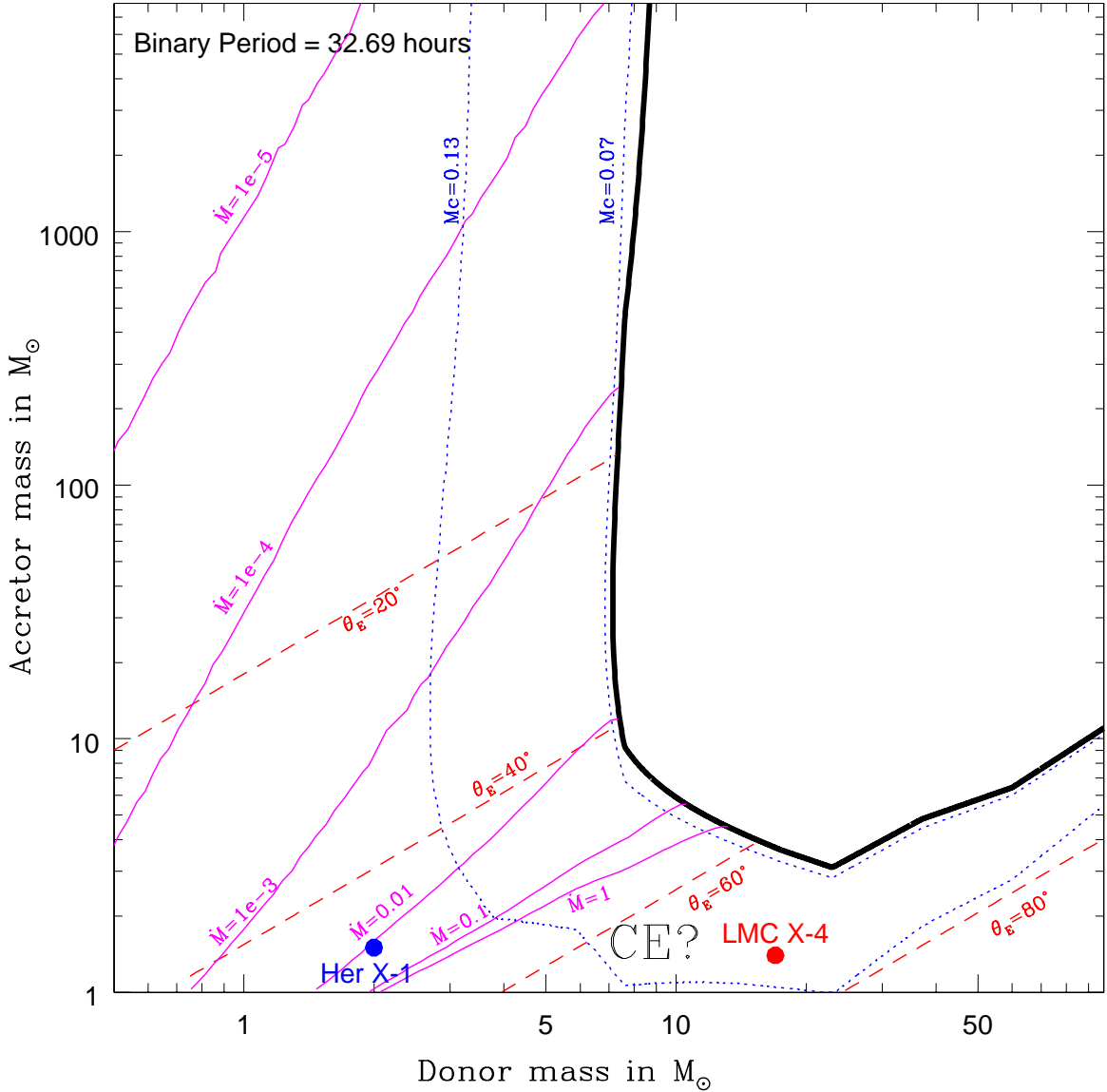}
\caption{Possible binary systems with the 32.69 hour period in the donor
mass-accretor mass phase plane. The thickest line stands for the viable
binaries with donors interpolated from the dwarf stars of all spectral types.
The region below and to the left of this line are populated by viable binaries
with different He core mass, and the dotted lines denote $M_c$=0.07 and 0.13
$M_\odot$.  The thin solid lines denote the accretion rates $\dot M_2$=1, 0.1,
0.01, $10^{-3}$, $10^{-4}$ and $10^{-5}$ $\bar L_X/c^2$, respectively. The
systems to the right of $\dot M_2$ = $\bar L_X/c^2$ may develop common
envelopes unless the accreted materials can be dissipated effectively. The
dashed lines denote the maximum eclipse full angles $\theta_E$ = $80^\circ$,
$60^\circ$, $40^\circ$ and $20^\circ$, respectively. }

\end{figure}

\clearpage

\begin{deluxetable}{lcrrrrcccccc}
\setlength{\tabcolsep}{0.025in}
\tabletypesize{\small}
\tablecaption{Individual observations for CXO J140336.0+541925\tablenotemark{a}}
\tablehead{
 \colhead{ObsID\tablenotemark{b}} & \colhead{MJD} & \colhead{ExpT} & \colhead{OAA} & 
 \colhead{$\sigma$} & \colhead{cnt} & \colhead{CR} & \colhead{$L_X$\tablenotemark{c}} & 
 \colhead{$n_H$} & \colhead{$\Gamma$} & \colhead{$\chi^2_\nu/dof$} & \colhead{Prob} 
}

\startdata

acis 934*  & 51629.01 & 99504 & 300.7 & 53.0 & 703 & 7.1 &1.9 & 0.8$\pm$0.3 & 2.7$\pm$0.2 & 1.133/21 & 0.30 \\
           &          &       &       &      &     &     &    & 0.16$\pm$0.42\tablenotemark{d}    & 0.41$\pm$0.03 & 2.156/21 & 0.0016\\
           &          &       &       &      &     &     &    & 3.3\tablenotemark{e}    & 3.3$\pm$0.2 & 1.123/14 & 0.33 \\
acis4731** & 53023.25 & 56963 & 306.5 & 19.6 & 150 & 2.6 &0.7 &           - &           - \\
acis5296   & 53025.44 &  3231 & 306.1 & 11.4 &  25 & 7.7 &2.1 &           - &           - \\
acis5297   & 53028.07 & 21964 & 306.4 & 25.4 & 158 & 7.2 &1.9 &           - &           - \\
acis5300** & 53071.40 & 52761 & 381.7 & 16.4 & 194 & 3.7 &1.0 &           - &           - \\
acis5309*  & 53078.07 & 71679 & 388.3 & 24.5 & 340 & 4.7 &1.3 & 2.7$\pm$0.7 & 3.0$\pm$0.4 & 1.739/9 & 0.07 \\
acis4732***& 53083.34 & 70691 & 388.4 & 24.9 & 315 & 4.5 &1.2 & 3.8$\pm$0.8 & 3.6$\pm$0.5 & 0.902/8 & 0.51 \\
acis5322***& 53128.32 & 65532 & 377.3 & 23.2 & 254 & 3.9 &1.1 &           - &           - \\
acis4733   & 53132.56 & 25132 & 373.5 &  2.1 &  24 & 0.9 &0.2 &           - &           - \\
acis5323*  & 53134.13 & 43161 & 372.2 & 26.6 & 304 & 6.7 &1.8 & 4.3$\pm$2.6 & 4.4$\pm$1.5 & 0.385/7 & 0.91 \\
acis5337   & 53191.78 & 10070 & 285.1 & 16.4 &  57 & 5.6 &1.5 &           - &           - \\
acis5338   & 53192.56 & 28934 & 285.3 & 14.4 &  57 & 2.0 &0.5 &           - &           - \\
acis5339   & 53193.56 & 14505 & 285.2 & 10.0 &  44 & 3.0 &0.8 &           - &           - \\
acis5340   & 53194.46 & 55116 & 285.4 & 21.4 & 107 & 1.9 &0.5 &           - &           - \\
acis4734   & 53197.07 & 35932 & 285.1 & 40.3 & 237 & 6.6 &1.8 &           - &           - \\
acis6114** & 53253.77 & 67052 & 116.0 & 48.5 & 312 & 4.7 &1.3 & 2.3$\pm$2.0 & 3.5$\pm$1.3 & 0.402/7 & 0.90 \\
acis6115*  & 53256.76 & 36204 & 115.8 & 49.8 & 232 & 6.4 &1.7 &           - &           - \\
acis6118   & 53259.64 & 11606 & 115.7 & 32.2 & 104 & 8.4 &2.3 &           - &           - \\
acis4735   & 53260.54 & 29148 & 116.0 & 13.9 &  45 & 1.6 &0.4 &           - &           - \\
acis4736***& 53310.78 & 78342 & 119.7 & 53.3 & 304 & 3.8 &1.0 & 2.9$\pm$0.2 & 3.7$\pm$1.3 & 0.364/7 & 0.92 \\
acis6152   & 53316.28 & 44656 & 119.3 & 15.7 &  66 & 1.5 &0.4 &           - &           - \\
acis6170***& 53361.05 & 48563 & 190.5 & 29.0 & 160 & 3.1 &0.8 &           - &           - \\
acis6175*  & 53363.69 & 41177 & 190.5 & 39.8 & 203 & 4.9 &1.3 &           - &           - \\
acis6169   & 53369.09 & 29753 & 190.9 & 53.7 & 222 & 7.5 &2.0 &           - &           - \\
acis4737   & 53371.60 & 22134 & 191.4 & 38.7 & 151 & 6.8 &1.8 &           - &           - \\

\enddata

\tablenotetext{a}{The columns are (1) ObsID, (2) MJD, (3) exposure time in
seconds, (4) Off-Axis Angle of the source in arcseconds, (5) wavdetect
detection significance, (6) net counts derived from aperture photometry, (7)
average net count rate in ct/ksec, 
%
%
(8) $L_X(0.3-8keV)$ in $10^{38}$ erg/sec, (9) absorbing column density in unit
of $10^{21}$ cm$^{-2}$ in the power law model, (10) the power law photon index,
(11) the reduced $\chi^2_\nu$ and degrees of freedom, and (12) the null
hypothesis probability. }

\tablenotetext{b}{Observations showing apparent variations are suffixed with a
'*'; an extra '*' is suffixed if the observation contains an apparent minimum;
a third '*' is suffixed if steep ingress and egress of a deep eclipse are present.  }

\tablenotetext{c}{1 ct/ks is converted to $2.7\times10^{37}$ erg/sec in 0.3-8.0
keV using the combined power-law fit, and a distance of 6.8 Mpc. The conversion
factor is 1 ct/ks = $2.4\times10^{37}$ erg/sec  if using the combined
multi-color disk fit. }

\tablenotetext{d}{The absorbed multi-color disk model is fitted in 0.3-8.0 keV,
with $n_H$ specified to be no less than the Galactic value of
$1.6\times10^{20}$ cm$^{-2}$. The second parameter (``$\Gamma$'') is the disk inner edge
temperature $T_{in}$. }

\tablenotetext{e}{The power-law model is fitted to the band 0.7-8 keV by
freezing $n_H=3.3\times10^{21}$ cm$^{-2}$ derived from the combined fit, to
exclude the spurious soft excess below 0.5 keV. }

\end{deluxetable}

\begin{deluxetable}{lrrrrr}
\tabletypesize{\small}
\tablecaption{Magnitudes for X7 counterpart candidates}
\tablehead{
 \colhead{Object} & \colhead{S/N} & \colhead{B} & \colhead{V} & \colhead{I} & \colhead{$H_\alpha$} \\
 \colhead{      } & \colhead{   } & \colhead{(mag)} & \colhead{(mag)} & \colhead{(mag)} & \colhead{(mag)} 
}

\startdata

A1 & 42.6 & 24.84 $\pm$ 0.04 & 24.96 $\pm$ 0.05 & 24.76 $\pm$ 0.06 & 24.39 $\pm$ 0.10 \\
A2 & 19.5 & 25.78 $\pm$ 0.07 & 26.16 $\pm$ 0.12 & 26.64 $\pm$ 0.27 \\
A3 & 6.2 & 27.13 $\pm$ 0.19 & 27.54 $\pm$ 0.38 & \nodata \\
B1 & 17.7 & 28.19 $\pm$ 0.45 & 27.80 $\pm$ 0.46 & 24.88 $\pm$ 0.06 & 24.81 $\pm$ 0.25 \\
B2 & 17.2 & 26.18 $\pm$ 0.09 & 26.14 $\pm$ 0.11 & 26.11 $\pm$ 0.16 \\
B3 & 16.1 & 26.19 $\pm$ 0.09 & 26.30 $\pm$ 0.13 & 26.32 $\pm$ 0.20 \\
C1 & 48.2 & 24.60 $\pm$ 0.03 & 24.65 $\pm$ 0.04 & 24.85 $\pm$ 0.06 & 25.32 $\pm$ 0.20 \\
C2 & 42.4 & 24.74 $\pm$ 0.03 & 24.86 $\pm$ 0.04 & 24.82 $\pm$ 0.12 \\
C3 & 30.2 & 26.89 $\pm$ 0.16 & 27.37 $\pm$ 0.34 & 24.05 $\pm$ 0.04 & 25.73 $\pm$ 0.28 \\
D1 & 78.0 & 23.76 $\pm$ 0.02 & 23.53 $\pm$ 0.02 & 23.23 $\pm$ 0.03 & 23.07 $\pm$ 0.04\tablenotemark{a} \\
D2 & 57.9 & 24.52 $\pm$ 0.04 & 24.55 $\pm$ 0.05 & 23.15 $\pm$ 0.02 \\
D3 & 41.5 & 25.65 $\pm$ 0.07 & 24.88 $\pm$ 0.05 & 23.84 $\pm$ 0.03 \\
D4 & 39.4 & 24.72 $\pm$ 0.04 & 24.73 $\pm$ 0.05 & 24.19 $\pm$ 0.05 \\
D5 & 24.8 & 25.80 $\pm$ 0.10 & 25.26 $\pm$ 0.08 & 24.30 $\pm$ 0.06 \\
E & 126.3 & 25.95 $\pm$ 0.07 & 23.98 $\pm$ 0.03 & 21.83 $\pm$ 0.01 & 22.86 $\pm$ 0.04 \\
F & 17.0 & 26.35 $\pm$ 0.09 & 26.12 $\pm$ 0.11 & 25.99 $\pm$ 0.15 \\
G & 9.5 & 26.41 $\pm$ 0.13 & 29.09 $\pm$ 0.29 & 27.68 $\pm$ 0.35 \\
H & 10.4 & 27.18 $\pm$ 0.18 & 26.41 $\pm$ 0.14 & 26.91 $\pm$ 0.33 \\
I & 9.3 & 26.91 $\pm$ 0.14 & 28.08 $\pm$ 0.22 & 28.53 $\pm$ 0.56 \\
J & 16.7 & 26.05 $\pm$ 0.07 & 26.42 $\pm$ 0.14 & 27.10 $\pm$ 0.38 \\

\enddata

\tablenotetext{a}{The listed $H_\alpha$ magnitude is for the counterpart D,
which was not resolved on the $H_\alpha$ image.}

\end{deluxetable}

\end{document}